\documentclass[a4paper,11pt]{article}%
\usepackage{graphicx}
\graphicspath{{Figures/}} 
\usepackage{amsmath}
\usepackage{amssymb}
\usepackage{subcaption}
\usepackage[colorlinks]{hyperref}
\usepackage{color}%
\usepackage{cite}
\usepackage{float}

\newcommand{\bra}{\begin{array}}
\newcommand{\era}{\end{array}}
\newcommand{\beq}{\begin{equation}}
\newcommand{\eeq}{\end{equation}}
\newcommand{\bqr}{\begin{eqnarray}}
\newcommand{\eqr}{\end{eqnarray}}
\newcommand{\rr}{\color{red}}

\def\BC{\bb C}
\def\_\BC{\bbi C}

\def\Tr {{\rm Tr}}

\def\( {\left(}
\def\) {\right)}
\def\no2 {{\textstyle{n\over 2}}}

\def\Tr {{\rm Tr}}

\newcommand{\om}{\omega}

\newcommand{\lb}{\label}

\begin{document}
\begin{titlepage}
\setcounter{page}{1}
\renewcommand{\thefootnote}{\fnsymbol{footnote}}
\begin{flushright}
\end{flushright}
\vspace{5mm}
\begin{center}
{\Large \bf { Magnetic Field Effect on Dynamics of Entanglement\\
for Time-dependent Harmonic Oscillator
}}

\vspace{5mm}

{\bf {{Radouan Hab-arrih}} }$^{a}$,
{\bf Ahmed Jellal\footnote{\sf a.jellal@ucd.ac.ma}}$^{a,b}$
and {\bf Abdeldjalil Merdaci}$^{b}$

\vspace{5mm}
{$^{a}$\em Laboratory of Theoretical Physics,  
Faculty of Sciences, Choua\"ib Doukkali University},\\
{\em PO Box 20, 24000 El Jadida, Morocco}

{$^{b}$\em Canadian Quantum  Research Center,  
204-3002 32 Ave Vernon, \\ BC V1T 2L7,  Canada}

{$^c$\em Department of Physics, Faculty of Sciences,
University of 20 août 1955-Skikda,\\
 Road El-Hadaeik, B.P. 26,
21000, Skikda, Algeria
}

\vspace{30mm}
\begin{abstract}
We investigate  the dynamics of entanglement, uncertainty and mixedness by solving time dependent Schr\"{o}dinger equation for 
two-dimensional harmonic oscillator with
time dependent frequency and coupling parameter subject to a
static magnetic field. 
We compute the purities (global/marginal) and then  calculate explicitly the linear entropy $S_{L}$   as well as  logarithmic 
negativity $\mathcal{N}$  using the symplectic parametrization of vacuum state. We introduce the spectral decomposition to diagonalize the marginal state and get the expression of von Neumann entropy $S_{von}$ and establish its link with
$S_{L}$. We use the Wigner formalism to derive the Heisenberg  uncertainties  and {\rr show their dependencies
 on both $S_{L}$ and the coupling parameters $\gamma_{i}$ $ (i=1,2)$ of the quadrature term $x_{i}p_{i}$.}
We graphically study the dynamics of the three features (entanglement, uncertainty, mixedness) and present the similar topology with respect to time. We show the effects of the magnetic field and  quenched values of $J(t)$ and $\omega_{2}(t)$  on these three dynamics, which lead eventually to control and handle them.

\vspace{3cm}

\end{abstract}
\end{center}
\end{titlepage}

\section{Introduction}
The entanglement, uncertainty and mixedness are three fundamental and remarkably features of
the quantum information. Indeed firstly, the entanglement is the most amazing property of
quantum mechanics, which expresses the "spooky" non-locality action \cite{R1}
between quantum states but it is still a physical reality \cite{R3}. With the 
entanglement one can describe a large type of physical phenomena and related issues \cite{R9}.
 The {von Neumann} entropy \cite{R1} of the reduced states or generally the {R\'{e}nyi} entropies \cite{R8,R30} can be used as suitable quantifiers of entanglement amount for pure bipartite states (i.e. $\mathcal{H}=\mathcal{H}_{A}\otimes\mathcal{H}_{B}$). However, for mixed states the quantification of non-local correlations is still an open problem \cite{R1,R31}. This   is due to the fact that there exist infinitely many pure state decomposition, which complicate such quantification. 
  Secondly, the uncertainty lies at the core of quantum physics and presents a key of the discrepancies between classical and quantum systems \cite{R32}. It can be understood mathematically as two observables, which are conjugate (i.e. connected by the {\rr Fourier} transform) and the trade-off between  their spreads can not be zero. In the early twentieth century, {Heisenberg} proposed in a seminal paper \cite{R33} showing the  variance-based uncertainty for position and momentum $\Delta x \Delta p\geq \frac{\hslash}{2}$. 
  This has been extended to generalized uncertainties, one of them is the {\rr Robertson-Schr\"{o}dinger} uncertainty $ \Delta_{A}\Delta_{B}\geq \frac{1}{2}\left[\vert\langle\lbrace A,B\rbrace\rangle-2\langle A\rangle\langle B\rangle\vert^{2}+\vert\langle\left[A,B\right]\rangle\vert^{2}\right]^{\frac{1}{2}}$, with  $A$ and $B$ are two arbitrary observables, 
  while the averages are with respect to  the quantum state $|{\Psi}\rangle$  encoding those fluctuations \cite{R34}. Recently  more general uncertainties based on entropy were introduced to be used  as entanglement witnesses  \cite{R35,R33}. 
  Thirdly, the mixedness is the loss of information caused by the preparation of states.
  Consequently,  the mixed state can be written as a convex linear combination of pure states $|{\Psi}\rangle=\sum_{i=1}^{n}p_{i}|\phi_{i}\rangle$ with 
  the conservation {\rr of the quantum probabilities, i.e.} $\sum_{i=1}^{n}p_{i}=1$. It is worthwhile noting that $(p_{i})_{i=1,\cdots, n}$ are classical probabilistic distributions, which is the point that distinguish the mixedness and superposition principle.
   We mention that different studies showed that the three {\rr 
    fundamental features, mentioned above,} are related  \cite{R8,R24,R36}.
  
The study of problems addressing to the coupled harmonic oscillators with time dependent frequencies  has of paramount importance in different scientific branches. This is due to its intrinsic mathematical interest and its power to model the behavior of systems in the vicinity of their equilibrium. Their time dependent Schr\"{o}dinger equation (TDSE) was solved exactly  using different methods \cite{R4,R5,R6,R7}. The solutions of TDSE were widely used to investigate the dynamics of entanglement \cite{R8,R9} and uncertainty \cite{R8} for bipartite systems. Many {\rr works} dealt with 
{\rr the Gaussian solutions 
(e.g. vacuum \cite{R8,R9}, thermal \cite{R30}, squeezed \cite{R24}, coherent $\cdots$)  and} have been considered as  prototypical continuum variables. {\rr This is quite   natural} because they can be created and assisted via linear optics. Recently, it was shown that the entanglement  can be assisted via temperature \cite{R30}, magnetic field \cite{R37}  and both  \cite{R38}, but  most of {\rr works} dealt with time independent potential.

We study {\rr the effect of magnetic field}  on the dynamics of entanglement 
for  
two-dimensional harmonic oscillator with
time-dependent frequency and coupling parameter.
We embark on the vacuum state
and determine the purity function, which allows us to quantify the entanglement
together with the degree of mixedness and {{quantum fluctuations}}. We show that
the magnetic field purifies the marginal states in decreasing 
the amount of quantum fluctuations and small values of the quenched frequencies  
increase them. In addition, 
using the Wigner function we derive the uncertainty relations in terms of the linear entropy telling us that 
its lower bound 
depends
on magnetic field and time. This is important feature to preserve the invariance of uncertainty relations with respect to phase space transformations during the
action of magnetic field and dynamics.

The present paper is organized as follows. In section \ref{S2}, we reduce the Hamiltonian into a diagonalized form using some transformations. In section \ref{S3}, we {\rr solve the associated TDSE}  and cast the eigen-spectrum of the time-dependent harmonic oscillator expressed with respect to original phase space coordinates. 
In section \ref{S4}, we compute the global  and  marginal purities, which {\rr will be used} to discuss the mixedness via linear entropy $S_{L}$. Then we obtain the decomposition of the marginal state $\rho^{A}_{(0,0)}$ {\rr that leading} to compute the {von Neumann} entropy $S_{von}$. 
In section \ref{S5}, we study the dynamics of entanglement via logarithmic negativity $\mathcal{N}$ and discuss its relationship to $S_{L}$. In section \ref{S6}, we calculate the fluctuations encoded in the vacuum state  using the {Heisenberg} uncertainty and discuss some limiting cases. In section \ref{S7}, we involve a realistic quenched model to follow the dynamics and study  the effects of magnetic field, quenched coupling $J_{f}$ 
{and frequency} $\omega_{2,f}$ on the dynamics of entanglement, uncertainty and mixedness. {\rr Finally}, we summarize our results.

\section{Diagonalized Hamiltonian\label{S2}}

{The machinery of 
the coupled harmonic oscillators is successfully used 
to investigate the effect of the rest of universe on the entropy and quantum fluctuations 
\cite{R77},  the Bogoliubov  transformation model of superconductivity  \cite{R88}, the magnetoresistance \cite{R99},  the cyclotron resonance \cite{R1010} and so on.
The
time-dependent harmonic oscillator has received much attention because of its applications in several areas
of physics \cite{Colegrave}, since the variable-mass and variable-frequency oscillators are
important in quantum optics  as well as in many other
fields. As for the 
time-dependent coupled harmonic
oscillators, 
they are the best candidate to describe several quantum mechanical problems. Among them, we cite 
the ion-laser interactions \cite{100}, quantized
fields propagating through dielectric media \cite{555}, shortcuts to adiabaticity 
\cite{666,777}, the Casimir effect \cite{888,999}. {\rr They can be used} as a
toy model to study aspects of the final stages of black-hole evaporation \cite{101} and their Schmidt modes in
the study of quantum entanglement \cite{111}.

Motivated by the above achievements, 
we consider 
a charged  particle in magnetic field
described by a two-dimensional harmonic oscillator
with
time-dependent frequency and coupling parameter in the presence of a
static perpendicular magnetic field. In this case,
the canonical momenta $\vec{p}=(p_{1},p_{2})$ will be
replaced by the conjugate momenta $(\pi_{1},\pi_{2})=\vec{p}-\frac{e}{c}\vec{A}$
where $\vec{A}$ is the potential vector such that  $\vec\nabla\wedge \vec A= \vec{B}$.
The corresponding Hamiltonian is
\begin{equation}
H_{0}=\frac{1}{2}\left(  \pi_{1}^{2}+\pi_{2}^{2}\right)  +\frac{1}{2}%
\omega_{1}^{2}(t)x_{1}^{2}+\frac{1}{2}\omega_{2}^{2} (t)x_{2}^{2}%
-J(t)x_{1}x_{2}%
\label{h1}
\end{equation}
where the  frequencies $\omega_{i}(t)$ ($i=1,2$) and coupling  $J(t)>0$ are three
parameters time-dependent.
To proceed further, we choose
the symmetric gauge $\vec A =\frac{B}{2} (x_2, -x_1)$ to obtain the conjugate momenta
\begin{equation}\lb{vpot}
\pi_{1}=p_{1}+\dfrac{eB}{2c}x_{2},\qquad\pi_{2}=p_{2}-\dfrac{eB}{2c}x_{1}%
\end{equation}
{\rr which can be  substituted into $H_{0}$ to get}
\begin{equation}
H_{1}=\dfrac{p_{1}^{2}+p_{2}^{2}}{2}+\frac{1}{2}\Omega_{1}^{2}(t)x_{1}%
^{2}+\frac{1}{2}\Omega_{2}^{2}(t)x_{2}^{2} -J(t)x_{1}x_{2}+\omega_{c}\left(  p_{1}x_{2}%
-p_{2}x_{1}\right)  %
\end{equation}
where we have set $\Omega_{i}^{2}(t)=\omega_{i}^{2}(t)+\omega_{c}^{2}$
and  
$\omega_{c}=\dfrac{eB}{2c}$ is the cyclotron frequency. 
It is clearly seen that
our system {\rr is now described by two coupled harmonic oscillators together with}
the angular momentum operator $L_z$. 
We will study the quantum information of  the 
time-dependent coupled harmonic oscillators 
by showing that $B$ plays a crucial role in the dynamics of entanglement encoded in the both continuum variable quantifiers von Neumann entropy $S_{von}$ and logarithmic negativity $E_{\mathcal{N}}$. This fact will due to strong dependence of the solutions $h_{i}(t,\omega_{c})$ (equations (\ref{eq82}-\ref{eq83})) of Ermakov equation  on the frequency $\omega_{c}$, which undergo
important amplitudes and frequency modulations with respect to $\omega_{c}$.
Note in passing that 
the Hamiltonian \eqref{h1}  
was previously established in the literature to study different issues.   For instance,  it is introduced to resolve the  both dynamics 
(quantum/classical) using two different approaches (unitary transformation/generating function)
and considering time-dependent quantities (mass, frequencies, magnetic field and coupling parameter) \cite{R26}.

To get rid of the operator $L_z$, 
we
introduce  a quantum canonical transformation from  
$\left(  x_{1},x_{2},p_{1},p_{2}\right)$ to new variables $\left(y_{1},y_{2},P_{1},P_{2}\right)$. This is 
\beq
\left(
\begin{matrix}
x_{1}\\
x_{2}%
\end{matrix}
\right)  =\left(
\begin{matrix}
\cos\phi\left(  t\right)   & \sin\phi\left(  t\right)  \\
-\sin\phi\left(  t\right)   & \cos\phi\left(  t\right)
\end{matrix}
\right)  \allowbreak\left(
\begin{matrix}
y_{1}\\
y_{2}%
\end{matrix}
\right)  ,\qquad\left(
\begin{matrix}
p_{1}\\
p_{2}%
\end{matrix}
\right)  =\left(
\begin{matrix}
\cos\phi\left(  t\right)   & \sin\phi\left(  t\right)  \\
-\sin\phi\left(  t\right)   & \cos\phi\left(  t\right)
\end{matrix}
\right)  \allowbreak\left(
\begin{matrix}
P_{1}\\
P_{2}%
\end{matrix}
\right)
\eeq
{\rr and we have}
 \cite{Goldstein}
\begin{align}
p_{1} &  =\frac{\partial\hat{F}_{2}\left(  x_{1},x_{2},P_{1},P_{2};t\right)
}{\partial x_{1}},\qquad p_{2}=\frac{\partial\hat{F}_{2}\left(  x_{1}%
,x_{2},P_{1},P_{2};t\right)  }{\partial x_{2}}\\
y_{1} &  =\frac{\partial\hat{F}_{2}\left(  x_{1},x_{2},P_{1},P_{2};t\right)
}{\partial P_{1}},\qquad y_{2}=\frac{\partial\hat{F}_{2}\left(  x_{1}%
,x_{2},P_{1},P_{2};t\right)  }{\partial P_{2}}%
\end{align}
where the {hermitian} generating function is given by
\begin{equation}
\hat{F}_{2}\left(  x_{1},x_{2},P_{1},P_{2};t\right)  =\frac{P_{1}x_{1}%
+x_{1}P_{1}+P_{2}x_{2}+x_{2}P_{2}}{2}\cos\phi\left(  t\right)  +\left(
P_{2}x_{1}-P_{1}x_{2}\right)  \sin\phi\left(  t\right)
\end{equation}
{\rr showing that}
\begin{equation}
\frac{\partial }{\partial
t} \hat{F}_{2}\left(  x_{1},x_{2},P_{1},P_{2};t\right) =\dot{\phi}\left(  t\right)  \left(  -P_{1}y_{2}+P_{2}y_{1}\right).
\end{equation}
According to \cite{Goldstein,R25}, the above quantum function can be used to express
the Hamiltonian 
into the new coordinate representation
as 
\begin{equation}
H_{2}=H_{1}+\frac{\partial\hat{F}_{2}\left(  x_{1},x_{2},P_{1},P_{2};t\right)
}{\partial t}=H_{1}-\dot{\phi}\left(  t\right)  \left(  P_{1}y_{2}-P_{2}%
y_{1}\right).
\end{equation}
 
It worthwhile to mention that
the method of canonical transformations (CT) and generating function (GF) proved to be a fruitful approach in treating quantum systems. Recall that  two systems described by the Hamiltonian's $H_0$ and $H_{1}$ are canonically equivalent if they are related by a canonical or equivalently unitary operation, which conserves all the quantum eigen-spectrum and the mean values of observables. 
The mapping  
$(\vec{\pi},x)\mapsto (\vec{p},x)$ is not a CT (i.e $[\pi_{1},\pi_{2}]=2i\omega_{c}$) but just a substitution  and we do not have any transformation on position representation $(x_{1},x_{2})$, then the transformation does not affect the Hamiltonian function $H_{0}(x_{1},x_{2},t)=H_{1}(x_{1},x_{2},t)$.
In the generating function theory \cite{Goldstein}, 
a generating function $F$ exists
and then the Hamiltonian  must verifies the condition 
$H_{1}(x_{1},x_{2})=H_{0}(x_{1},x_{2})+ \frac{\partial F}{\partial t}$, obviously   
 $\frac{\partial F}{\partial t}=0$ for time-independent systems, more clarification can be found in \cite{Goldstein}. 
Now,
imposing the condition $\dot{\phi}\left(  t\right)  =\omega_{c}$ 
we obtain {\rr a linear equation in time}
\begin{equation}
\phi\left(  t\right)  =\omega_{c}t+\theta \lb{phit} 
\end{equation}%
and the angle $\theta$ is a constant of integration. In this case, 
$H_2$ becomes 
\begin{equation}
H_{2}=\frac{1}{2}\left(  P_{1}^{2}+P_{2}^{2}\right)  +\frac{1}{2}\sigma
_{1}^{2}(t,B)y_{1}^{2}+\frac{1}{2}\sigma_{2}^{2}(t,B)y_{2}^{2}%
+\sigma_{3}(t,B)y_{1}y_{2}%
\label{ham11}
\end{equation}
{\rr with} $\sigma_{j}^{2}=\varsigma^{2}_{j}+\omega_{c}^{2}$ $ (j=1,2) $ and
{\rr
\begin{align}
\varsigma_{1}^{2}(t,B) &  =\omega_{1}^{2}(t)\cos^{2} \phi(t)
+\omega_{2}^{2}(t)\sin^{2} \phi(t) +J(t)\sin 2\phi(t)  \\
\varsigma_{2}^{2}(t,B) &  =\omega_{1}^{2}(t)\sin^{2}\phi(t)  +\omega_{2}^{2}(t)\cos^{2}\phi(t)  -J(t)\sin 2\phi(t)  \\
\sigma_{3}(t,B) &  =\tfrac{\omega_{1}^{2}(t)-\omega_{2}^{2}(t)}%
{2}\sin 2\phi(t)  -J(t)\cos 2\phi(t).\lb{144}
\end{align}}At this stage, let us look for  {\rr the appropriate  rotation of  angle $\phi(t)$} that leads to the uncoupled harmonic oscillators. 
To this end, we use a method extremely employed in different context to
deal with many issues. For instance by selecting the appropriate rotation, Han {\it et al.} \cite{Han1999}
showed 
that the coupled harmonic oscillator can be diagonalized and would yield an increased entropy in the observable oscillator,
providing a clarification of Feynman's rest of the universe. Subsequently,
the same technique has been used in generalizing the  Han {\it et al.} \cite{Han1999} work  to 
the non-commutative 
space \cite{Jellal1}, studying the entanglement of coupled harmonic oscillators 
\cite{Jellal2, R8, Jellal3, Jellal4} and proposing 
an alternative approach to exact wave functions for time-dependent coupled oscillator model of charged particle in
variable magnetic field \cite{R26}.
Similarly, we can choose the rotation 
{\rr
\begin{align}
\tan2 \phi(t)= \frac{2J(t)}{\om_1^2(t)-\om_2^2(t)}\label{e15}
\end{align}}
to ensures
the solvability of the Hamiltonian describing our system. We emphasis that  \eqref{e15} is a
necessary and sufficient condition (NSC), which holds for all time 
and corresponds to $\sigma_{3}(t, B)=0$ in \eqref{144}. Since it depends only on the original parameters of the problem $(\omega_{i}(t),J(t))$ then  NSC leads directly to an uncoupled {\rr Hamiltonian \eqref{ham11} in the $(y_{1},y_{2})$-representation}.
In our knowledge, this is the unique way to get a diagonalized Hamiltonian
and therefore derive the solutions of the energy spectrum.
%
%
Now, 
we have some remarks in order. Firstly, \eqref{e15} is actually connecting 
the magnetic field $B$  to frequencies $\omega_{i}(t)$ and coupling parameter $J(t)$.
{\rr 
Secondly,}
if the coupling parameter $J(t)$ is switched off then there is no rotation meaning that $\phi(t)=0$ and automatically both $B=0$ and $\theta$ are null. 
 {\rr Thirdly for zero time, we recover the rotation with an angle used} previously in literature \cite{Han1999, Jellal1, Jellal2}, which is
\beq
 \tan2\theta=\frac{2J(0)}{\om_1^2(0)-\om_2^2(0)}
\label{e155}.
\eeq
Consequently,
the Hamiltonian becomes diagonal}
\begin{equation}
H_{2}=\frac{1}{2}\left(  P_{1}^{2}+P_{2}^{2}\right)  +\frac{1}{2}\sigma
_{1}^{2}(t,B)y_{1}^{2}+\frac{1}{2}\sigma_{2}^{2}(t,B
)y_{2}^{2}\label{Eq17}
\end{equation}
which will be solved to determine the eigenvalues and eigenfunctions
using some techniques  involving
 time dependent frequencies.

\section{Exact wavefunctions\label{S3}}

{\rr We recall that 
the time dependent Schr\"odinger equation (TDSE) for any single harmonic oscillator having a frequency 
time-dependent can be 
written as}
\begin{equation}
-\frac{1}{2}\left[  \frac{\partial^{2}}{\partial y_{1}^{2}}-\sigma_{1}%
^{2}(t,B)y_{1}^{2}\right]  \Psi(y_{1},t)=i\frac{\partial\Psi
}{\partial t}%
\end{equation}
that  can be 
exactly solved \cite{R6} using 
theory of invariants 
\cite{R4,R5} or  transformation group techniques \cite{R7}. The solutions 
are
\begin{equation}
\Psi_{n}\left(  y_{1},t\right)  =e^{-i\int_{0}^{t}E_{n}\frac{dt^{\prime}%
}{h_{1}^{2}(t^{\prime})}}e^{\frac{i}{2}\left(  \frac{\dot{h}_{1}}{h_{1}%
}\right)  y_{1}^{2}}\Phi_{n}\left(  \frac{y_{1}}{h_{1}}\right), \qquad n\in \mathbb{N}
\end{equation}
such that the eigenstates $\Phi_{n}$ and eigenvalues $E_n$
are given by
\begin{align}
&  \Phi_{n}\left(\frac{y_{1}}{h_{1}}\right)=\dfrac{1}{\sqrt{2^{n}n!}}\left(
\dfrac{\tilde{\sigma}_{1}(t,B)}{\pi}\right)  ^{\frac{1}{4}}%
H_{n}\left(  \sqrt{\tilde{\sigma}_{1}(t,B)}y_{1}\right)
e^{\frac{-\tilde{\sigma}_{1}(t,B)}{2}y_{1}^{2}}\\
&  E_{n}= \sigma_{1}(0,B)\left(n+\frac{1}{2}\right)
\end{align}
where  $H_{n}(\epsilon)$ are Hermite polynomials,
$\tilde{\sigma}_{1}(t,B)=\frac{\sigma_{1}(0,B)}{h_{1}%
^{2}(t)}$ with the function $h_{1}(t)$ satisfies the {Ermakov} equation
\begin{equation}
\ddot{h}_{1}+\sigma_{1}^{2}(t,B)h_{1}=\dfrac{\sigma_{1}^{2}%
(0,B)}{h_{1}^{3}}%
\end{equation}
and two initial conditions $h_{1}(0)=1$, $\dot{h}_{1}(0)=0$. 
Therefore,
the wavefunctions associated to the Hamiltonian \eqref{Eq17} are
\begin{eqnarray}
\Psi_{n,m}(y_{1},y_{2}:t) &  =&\dfrac{1}{\sqrt{2^{n+m}n!m!}}\left(
\dfrac{\tilde{\sigma}_{1}\tilde{\sigma}_{2}}{\pi^{2}}\right)  ^{\frac{1}{4}%
}e^{-i\left(  \int_{0}^{t}E_{n}\frac{dt^{\prime}}{h_{1}^{2}(t^{\prime})}%
)+\int_{0}^{t}E_{m}\frac{dt^{\prime}}{h_{2}^{2}(t^{\prime})}\right)  }\nonumber\\
&&
e^{-\frac{1}{2}\left(  \tilde{\sigma}_{1}y_{1}^{2}+\tilde{\sigma}_{2}y_{2}%
^{2}\right)  } 
  \ e^{\frac{i}{2}\left[  \left(  \frac{\dot{h}_{1}}{h_{1}}y_{1}%
^{2}+\frac{\dot{h}_{2}}{h_{2}}y_{2}^{2}\right)  \right]  }H_{n}\left(
\sqrt{\tilde{\sigma}_{1}}y_{1}\right)  H_{m}\left(  \sqrt{\tilde{\sigma}_{2}%
}y_{2}\right)
\end{eqnarray}
and 
$\tilde{\sigma}_{2}(t,B)=\frac{\sigma_{2}(0,B)}{h_{2}%
^{2}(t)}$.
In terms of 
the original coordinates we have
\begin{eqnarray}
\Psi_{n,m}(x_{1},x_{2}:t) &  =&\dfrac{1}{\sqrt{2^{n+m}n!m!}}\left(
\frac{\tilde{\sigma}_{1}\tilde{\sigma}_{2}}%
{\pi^{2}}\right)  ^{\frac{1}{4}}e^{-i\left(  \int_{0}^{t}E_{n}\frac
{dt^{\prime}}{h_{1}^{2}(t^{\prime})}+\int_{0}^{t}E_{m}\frac{dt^{\prime}}%
{h_{2}^{2}(t^{\prime})}\right)  }\label{e23} \nonumber\\
&&  e^{-\frac{1}{2}\rho_{1}\left(  \cos(\phi)x_{1}-\sin(\phi
)x_{2}\right)  ^{2}-\frac{1}{2}\rho_{2}\left(  \sin(\phi)x_{1}+\cos(\phi
)x_{2}\right)  ^{2}}\\
&&   H_{n}\left(  \sqrt{\tilde{\sigma}_{1}}\left(  \cos(\phi)x_{1}%
-\sin(\phi)x_{2}\right)  \right)  H_{m}\left(  \sqrt{\tilde{\sigma}_{2}%
}\left(  \sin(\phi)x_{1}+\cos(\phi)x_{2}\right)  \right)  \nonumber
\end{eqnarray}
where 
$\rho_{j}(t,B)=\tilde{\sigma}_{j}(t,B)-i\frac{\dot{h}%
_{j}}{h_{j}}(t)$ and $j=1,2$. 
%
The corresponding vacuum state can be written in compact form as
\begin{eqnarray}
\Psi_{0,0}(x_{1},x_{2}:t)  =\left(  \frac{\tilde{\sigma}_{1}\tilde{\sigma}_{2}}{\pi^{2}}\right)  ^{\frac{1}{4}%
}e^{-\frac{i}{2}\left(  \sigma_{1}(0,B)\int_{0}^{t}\frac{dt^{\prime}%
}{h_{1}^{2}(t^{\prime})}+\sigma_{2}(0,B)\int_{0}^{t}\frac{dt^{\prime
}}{h_{2}^{2}(t^{\prime})}\right)  }
  e^{-\frac{1}{2}A_{1}x_{1}^{2}-\frac{1}{2}A_{2}x_{2}^{2}+A_{12}%
x_{1}x_{2}}
\end{eqnarray}
where have defined  three  time dependent parameters
\begin{eqnarray}
&& A_{1}(t,B)=\rho_{1}\cos^{2}\phi(t)+\rho_{2}\sin^{2}\phi(t
)\\
&& A_{2}(t,B)=\rho_{2}\cos^{2}\phi(t)+\rho_{1}\sin^{2}%
\phi(t) \\
&& A_{12}(t,B)=\sin\phi(t)\cos\phi(t)(\rho_{1}-\rho_{2})
\end{eqnarray}
showing the
identity
$\Re(A_{1})\Re(A_{2})-\Re^{2}(A_{12})=\tilde{\sigma}_{1}\tilde{\sigma}_{2}$
and $\Re(\xi)$ denotes the real part of $\xi\in\mathbb{C}$. Next, we will see
how to use the above results to discuss different issues related to the quantification
of information.

\section{Mixedness and entanglement \label{S4}} 

To discuss entanglement and mixedness of the vacuum state, we  first introduce 
density matrix, which is nothing but the product
\begin{align}
\rho_{0,0}^{AB}(x_{1},x_{2}:x_{1}^{'},x_{2}^{'}:t) & =\Psi
_{0,0}^{AB}(x_{1},x_{2}:t)\ \Psi_{0,0}^{\ast AB}(x_{1}',x_{2}':t)\\
&  =\left(  \frac{\tilde{\sigma}_{1}\tilde{\sigma}_{2}}{\pi^{2}}\right)
^{\frac{1}{2}}e^{-\frac{1}{2}(A_{1}x_{1}^{2}+A_{1}^{\ast}x_{1}^{^{\prime}%
2}+A_{2}x_{2}^{2}+A_{2}^{\ast}x_{2}^{^{\prime}2})+A_{12}x_{1}x_{2}%
+A_{12}^{\ast}x_{1}^{^{\prime}}x_{2}^{^{\prime}}}
\end{align}
{\rr where}
 the {global state is pure}
\begin{eqnarray}
P\left[\rho_{(0,0)}^{AB}\right]=\Tr\left[\left(\rho^{AB}_{(0,0)}\right)^{2}\right]
=1.
\end{eqnarray}
 For the reduced states $(\rho_{(0,0)}^{A},\rho_{(0,0)}^{B})$, we consider  one accessible of the two harmonic oscillators  says A and the other  remains  inaccessible. Then 
the reduced
density matrix associated to A is
{\rr
\begin{eqnarray}
\rho_{0,0}^{A}(x_{1},x_{1}^{^{\prime}}:t) &=&\Tr_{B}\rho_{0,0}^{AB}=\int dx_{2}%
\rho_{0,0}^{AB}(x_{1},x_{2},x_{1}^{^{\prime}},x_{2}:t)
\\
&=&\left(  \frac{\tilde{\sigma}%
_{1}\tilde{\sigma}_{2}}{\pi}\right)  ^{\frac{1}{2}}\left(  \frac{1}{\Re(A_{2}%
)}\right)  ^{\frac{1}{2}}e^{-\frac{1}{2}D_{1}x_{1}^{2}-\frac{1}{2}D_{2}%
x_{1}^{^{\prime}2}+\frac{1}{2}D_{12}x_{1}x_{1}^{^{\prime}}}%
\end{eqnarray}} where we have defined
\begin{align}
D_{1}=A_{1}-\frac{A_{12}^{2}}{2\Re(A_{2})}, \qquad
  D_{2}=A_{1}^{\ast}-\frac{A_{12}^{\ast2}}{2\Re(A_{2})}=D_{1}^{\ast}, \qquad
  D_{12}=\frac{|A_{12}|^{2}}{\Re(A_{2})} 
\end{align}
showing  the relation
$D_{1}+D_{2}-D_{12}=\frac{2\tilde{\sigma}_{1}\tilde{\sigma}_{2}}{\Re(A_{2})}$.

On the other hand, to shed light on the degree of mixedness in our system one
can compute the linear entropy 
\begin{equation}
S_{L}=1-\Tr\left[  \left(  \rho_{0,0}^{A}\right)  ^{2}\right]
\end{equation}
such that the trace is given by
\begin{align}
\Tr\left[  \left(  \rho_{0,0}^{A}\right)  ^{2}\right]   &  =\int
dxdx^{^{\prime}}\rho_{0,0}^{A}(x,x^{^{\prime}}:t)\ \rho_{0,0}%
^{A}(x^{^{\prime}},x:t)\nonumber\\
&=\left(  \dfrac{\tilde{\sigma}_{1}\tilde{\sigma}_{2}}{\tilde{\sigma}%
_{1}\tilde{\sigma}_{2}+|A_{12}|^{2}}\right)  ^{\frac{1}{2}}\leq1.\label{38}
\end{align}
It is clear that we have 
$\Tr\left[  \left(  \rho_{0,0}^{A}\right)  ^{2}\right]=\Tr\left[  \left(  \rho_{0,0}^{B}\right)  ^{2}\right]$ telling us
that the global state is  symmetric \cite{R24}.
Note that by requiring the limit $\omega_{c}\longrightarrow0$ we recover the result obtained in \cite{R8}. Thus, we conclude that the state $\rho$ is totally
mixed ($S_{L}=1$) if one of the frequencies vanishes, while it is a pure
state ($S_{L}=0$) if $A_{12}=0$, which is equivalent to the isotropic
oscillators $\left(\tilde{\sigma}_{1},\frac{\dot{h}_{1}}{h_{1}}\right)=\left(\tilde{\sigma
}_{2},\frac{\dot{h}_{2}}{h_{2}}\right)$ and the angle $\phi=k\pi,\frac{\pi}{2}+k\pi$ with
$k\in\mathbb{Z}$.

As outlined before our purpose here is to
measure the {von Neumann} entropy {{$S_{von}$}} of entanglement  and before doing
we 
compute the spectrum  $Sp\left(  \rho_{0,0}%
^{A}\right)  =\{p_{n},n\in\mathbb{N}\}\subset\lbrack0,1]$
of $\rho_{0,0}^{A}$, which is solution of the spectral equation
\begin{equation}
\int dx^{^{\prime}}\rho_{0,0}^{A}(x,x^{^{\prime}}:t)\ \chi_{n}(x^{^{\prime}%
},t)=p_{n}(t)\chi_{n}(x,t)\label{esp}
\end{equation}
and the computation gives
the normalized eigenfunctions as well as eigenvalues \cite{R9,R8,R19}
\begin{eqnarray}
&& \chi_{n}(x,t)=\frac{1}{\sqrt{2^{n}n!}}\left(  \frac{\kappa}{\pi}\right)
^{\frac{1}{4}}\ H_{n}\left(  \sqrt{\kappa}x\right)  e^{-\frac{\kappa}{2}%
x^{2}+i\alpha_{2}x^{2}}%
\\
&&
p_{n}(t)=(1-\gamma(t))\gamma^{n}(t)<1
\end{eqnarray}
where we have set the quantities
\begin{align}
&  \kappa(t,B)=2\left[  \alpha_{1}(\alpha_{1}+2\alpha_{3})\right]
^{\frac{1}{2}}\\
& \gamma(t,B)=\dfrac{\alpha_{3}}{(\alpha
_{1}+\alpha_{3})+\frac{\kappa}{2}}<1\\
&  \alpha_{1}(t,B) 
=\frac{\tilde{\sigma}_{1}\tilde{\sigma}_{2}}{2\Re(A_{2})}\\
&  \alpha_{2}(t,B)=\frac{\frac{\dot{h}_{1}}{h_{1}}\tilde{\sigma}%
_{2}\cos^{2}\phi(t)+\frac{\dot{h}_{2}}{h_{2}}\tilde{\sigma}_{1}\sin
^{2}\phi(t)}{2\left(  \tilde{\sigma}_{1}\sin^{2}\phi(t)+\tilde{\sigma}_{2}
\cos^{2}\phi(t)\right)  }\\
&  \alpha_{3}(t,B) 
=\frac{1}{4}\dfrac{\sin^{2}%
\phi(t)\cos^{2}\phi(t)\left[  (\tilde{\sigma}_{1}-\tilde{\sigma}_{2} 
)^{2}+\left(  \frac{\dot{h}_{1}}{h_{1}}-\frac{\dot{h}_{2}}{h_{2}}\right)
^{2}\right]  }{\tilde{\sigma}_{1}\sin^{2}\phi(t)+\tilde{\sigma}_{2}\cos
^{2}\phi(t)}
\end{align}
giving rise to  algebraic decomposition
$
D_{1,2}=\alpha_{1}+\alpha_{3}\mp i\alpha_{2}. 
$ 
Using the spectral decomposition theorem, one can easily show 
\begin{equation}
\rho_{0,0}^{A}(x,x^{\prime}:t)=\sum_{n=0}^{\infty}p_{n}(t)\varrho
_{n}(x,x^{\prime}:t)
\end{equation}
such that $\varrho_{n}(x,x^{\prime}:t)=\chi_{n}(x,t)\
\chi_{n}^{\ast}(x^{\prime},t)$.

As far as the probability distribution $\mathcal{P}=(p_{i})_{i=0,1 \cdots}$
is concerned, we
propose to compute {{$S_{von}$}}  and therefore quantify the entanglement. The analytic expression of {{$S_{von}$}} for a
bipartite state was shown originally in \cite{R12} 
\begin{equation}
S_{von}^{A}(\rho_{0,0}^{A})=-\ln(1-\gamma)-\frac{\gamma}{1-\gamma}\ln\gamma.
\end{equation}
For the limiting case where the frequencies are time-independent then we have
$h_{i}(t)=1$ and $\dot{h}_{i}(t)=0$. If $\phi=\frac{\pi}{4}+k\pi
,\,k\in\mathbb{Z}$ then 
\begin{eqnarray}
& \alpha_{1}\longrightarrow\frac{\sigma_{1}(0,B)\ \sigma_{2}(0,B)}{\sigma_{1}(0,B)+\sigma
_{2}(0,B)}, \qquad 
 \alpha_{3}\longrightarrow\frac{1}{8}\frac{\left[
\sigma_{1}(0,B)-\sigma_{2}(0,B)\right]  ^{2}}{\sigma
_{1}(0,B)+\sigma_{2}(0,B)}
\\
&
\gamma\longrightarrow\frac{\left[  \sigma_{1}(0,B)-\sigma_{2}%
(0,B)\right]  ^{2}}{\left[  \sigma_{1}(0,B)+\sigma
_{2}(0,B)\right]  ^{2}+4\sqrt{\sigma_{1}(0,B)\sigma
_{2}(0,B)}\left[  \sigma_{1}(0,B)+\sigma_{2}(0,B)+\sqrt{\sigma_{1}(0,B)
\sigma_{2}(0,B)}\right]  }.
\end{eqnarray}
At this {\rr level} we have some comments in order. Indeed,
if now the system oscillates in isotropic regime, i.e. $\sigma_{1}(0,B)\longrightarrow\sigma_{2}(0,B)$, then $\gamma\longrightarrow 0^{+}$ and
eventually $S_{von}\longrightarrow 0$, which means that the two vacuum states are
not entangled. It is clearly seen that if one of the frequencies approaches
$0$, $\gamma\longrightarrow1^{-}$ then $S_{von}\longrightarrow+\infty$. To generalize
this issue one can notice that in isotropic regime the {Ermakov} solutions
$h_{i}$ are equal and then we obtain $\alpha_{3}\longrightarrow 0$ giving rise to $S_{von}\longrightarrow 0$ {\rr showing that}
 the oscillators are separable. It is {{interesting}}  to investigate the
crucial role can be {{played}} by  magnetic field on the dynamics of entanglement, {\rr through the cyclotron frequency $\omega_{c}$}. In fact, for
$\omega_{c}\longrightarrow 0$ and $\theta\longrightarrow k\pi,\,k\in\mathbb{Z}$ we obtain
$\gamma\longrightarrow 0$ and then the subsystems are not entangled, which is obvious
because they are decoupled (i.e. $J\longrightarrow 0$) see (\ref{e15}).
%
The dynamics of entanglement can be {{analyzed}} directly from the
mixedness, because after a simple calculation we find
\begin{equation}
S_{von}=-\ln\left(  \frac{1-S_{L}}{1+S_{L}}\right)  -\frac{2S_{L}}{1-S_{L}}%
\ln\left(  \frac{2S_{L}}{1+S_{L}}\right).
\end{equation}
Thus for a linear entropy $S_{L}\longrightarrow 1$, the state is maximally mixed then $S_{von}%
\longrightarrow+\infty$, while for $S_{L}\longrightarrow 0$ we get $S_{von}%
\longrightarrow 0$. These limiting cases show  that $S_{von}$ increases as the mixedness amount increases.
 
\section{
Logarithmic negativity \label{S5} }

Recall that the logarithmic negativity $\mathcal{N}$ is a convenient measure of continuum
variables (CV) of entanglement \cite{{R22}, {R23}}. We notice that our time dependent ground states (TDGS) are the prototypical
quantum states and  their covariant matrix (CM) $V$ is given by \cite{R29}
\begin{eqnarray}
&& V_{ij}=\frac{1}{2}\left\langle \left\lbrace Q_{i}%
,Q_{j}\right\rbrace \right\rangle -\left\langle Q_{i}\right\rangle
\left\langle Q_{j}\right\rangle
\end{eqnarray}
such that $Q=\left(x_{1},p_{1},x_{2},p_{2}\right)\in \mathbb{R}^{4}$  is a vector of quadrature  phase satisfying the symplectic canonical commutation relations $[Q_{i},Q_{j}]=2i\mathcal{O}_{ij}$, with  the symplectic form $\mathcal{O}=\oplus_{i=1}^{2}\theta$, $\theta=\delta_{ij-1}-\delta_{ij+1}$, and $i,j=1,2$. Using our results
and making symplectic transformation, we show that CM takes the standard form  
\begin{equation}
V_{sf}=\left(
\begin{array}
[c]{cccc}%
\alpha(t,B) & 0 & \sqrt{\alpha^{2}(t,B)-1} & 0\\
0 & \alpha(t,B) & 0 & -\sqrt{\alpha^{2}(t,B)-1}\\
\sqrt{\alpha^{2}(t,B)-1} & 0 & \alpha(t,B) & 0\\
0 & -\sqrt{\alpha^{2}(t,B)-1} & 0 & \alpha(t,B)
\end{array}
\right)
\end{equation}
where $\alpha(t,B)$ is {\rr given by} 
\begin{equation}
 \alpha(t,B)=1+\frac{\sin^{2} {{2\phi(t)}}\left[\left(\frac{\dot{h}_{1}}{h_{1}}-\frac{\dot{h}_{2}}{h_{2}}\right)^{2}+\left(\tilde{\sigma}_{1}-\tilde{\sigma}_{2}\right)^{2}\right]}{4\tilde{\sigma}_{1}\tilde{\sigma}_{2}}.
\end{equation}
 The logarithmic negativity is defined by \cite{R24}
 \begin{equation}
 E_{\mathcal{N}}=\max\lbrace 0,-\log\tilde{\epsilon}_{-}\rbrace
 \end{equation}
 and {{$ \tilde{\epsilon}_{-}$ is the symplectic eigenvalue of $\tilde{V}_{sf}$ (the partial transpose of  
 $V_{sf}$)}}. After a simple calculation one can show that $\mathcal{N}$ 
 can be written in terms of the linear entropy $S_{L}$ and eventually make the relation between the loss of information encoded in $S_{L}$ and the amount of quantum correlations encoded in $\mathcal{N}$. Thus, we have
\begin{equation}
\mathcal{N}=-\log\left(\frac{1-S_{L}\left(2-S_{L}\right)^{\frac{1}{2}}}{1+S_{L}\left(2-S_{L}\right)^{\frac{1}{2}}}\right)^\frac{1}{2}.
\end{equation} 
 Note that for 
 $S_{L}\longrightarrow 1$, the state is maximally mixed then $\mathcal{N}%
\longrightarrow+\infty$, while for $S_{L}\longrightarrow 0$ we get $\mathcal{N}%
\longrightarrow 0$. 
Thus, we conclude 
that the two quantities $\mathcal{N}$ and $S_{von}$ present the same asymptotic behavior with respect to $S_{L}$. 


\section{Quantum fluctuations\label{S6}}

We study the quantum fluctuations for our system using the {Wigner}
formalism, in which the {Wigner} distribution $\mathcal{W}_{0,0}(x_{1},x_{2}:p_{1},p_{2}:t)$ associated to the vacuum state  is 
\begin{align}\lb{600}
\mathcal{W}_{0,0} (x_{1},x_{2}:p_{1},p_{2}:t)  
&  :=\frac{1}{\pi^{2}}\int
dq_{1} dq_{2}\ \psi^{\ast}_{0,0}(x_{1}+q_{1},x_{2}+q_{2}:t)\ \psi_{0,0}
(x_{1}-q_{1},x_{2}-q_{2}:t)\ 
e^{-2i(p_{1}q_{1}+p_{2}q_{2})}\nonumber\\
&  =  \frac{1}{\pi^{2}} e^{-\eta_{1}x_{1}^{2}-\eta_{2}x_{2}^{2}-\beta
_{1}p_{1}^{2}-\beta_{2}p_{2}^{2}+2\eta_{12}x_{1}x_{2}+2\beta_{12}p_{1}p_{2}
+2\delta_{1}x_{1}p_{2}+2\delta_{2}x_{2}p_{1}+2\gamma_{1}%
x_{1}p_{1}+2\gamma_{2}x_{2}p_{2}}
\end{align}
and the involved  quantities are
\begin{eqnarray}
&&  \eta_{1}(t,B)=\frac{1}{\tilde{\sigma}_{1}\tilde{\sigma}_{2}%
}\left(  \tilde{\sigma}_{1}\tilde{\sigma}_{2}\Re(A_{1})+\tilde{\sigma}%
_{2}\left(  \frac{\dot{h}_{1}}{h_{1}}\right)  ^{2}\cos^{2}(\phi)+\tilde
{\sigma}_{1}\left(  \frac{\dot{h}_{2}}{h_{2}}\right)  ^{2}\sin^{2}%
(\phi)\right) \\
&&  \eta_{2}(t,B)=\frac{1}{\tilde{\sigma}_{1}\tilde{\sigma}_{2}%
}\left(  \tilde{\sigma}_{1}\tilde{\sigma}_{2}\Re(A_{2})+\tilde{\sigma}%
_{1}\left(  \frac{\dot{h}_{2}}{h_{2}}\right)  ^{2}\cos^{2}(\phi)+\tilde
{\sigma}_{2}\left(  \frac{\dot{h}_{1}}{h_{1}}\right)  ^{2}\sin^{2}%
(\phi)\right) \\
&&  \eta_{12}(t,B)=\frac{\sin2\phi}{2\tilde{\sigma}_{1}\tilde
{\sigma}_{2}}\left(  \tilde{\sigma}_{1}\tilde{\sigma}_{2}(\tilde{\sigma}%
_{1}-\tilde{\sigma}_{2})+\tilde{\sigma}_{2}\left(  \frac{\dot{h}_{1}}{h_{1}%
}\right)  ^{2}-\tilde{\sigma}_{1}\left(  \frac{\dot{h}_{2}}{h_{2}}\right)
^{2}\right) \\
&&  \beta_{1}(t,B)=\frac{\Re(A_{2})}{\tilde{\sigma}_{1}\tilde{\sigma
}_{2}},\qquad\beta_{2}(t,B)=\frac{\Re(A_{1})}{\tilde{\sigma}_{1}%
\tilde{\sigma}_{2}},\qquad\beta_{12}(t,B)=-\frac{\sin2\phi}%
{2\tilde{\sigma}_{1}\tilde{\sigma}_{2}}(\tilde{\sigma}_{1}-\tilde{\sigma}%
_{2})\\
&&  \delta_{1}(t,B)=\frac{\sin2\phi}{2\tilde{\sigma}_{1}%
\tilde{\sigma}_{2}}\left(  \frac{\dot{h}_{1}}{h_{1}}\tilde{\sigma}_{2}%
-\frac{\dot{h}_{2}}{h_{2}}\tilde{\sigma}_{1}\right)=  \delta
_{2}(t,B) 
\\
&&  \gamma_{1}(t,B)=-\frac{1}{\tilde{\sigma}_{1}\tilde{\sigma}_{2}%
}\left(  \tilde{\sigma}_{2}\frac{\dot{h_{1}}}{h_{1}}\cos^{2}\phi
+\tilde{\sigma}_{1}\frac{\dot{h_{2}}}{h_{2}}\sin^{2}\phi\right)\lb{ga1}\\
&&
\gamma_{2}(t,B)=-\frac{1}{\tilde{\sigma}_{1}\tilde{\sigma}%
_{2}}\left(  \tilde{\sigma}_{1}\frac{\dot{h_{2}}}{h_{2}}\cos^{2}\phi
+\tilde{\sigma}_{2}\frac{\dot{h_{1}}}{h_{1}}\sin^{2}\phi\right).\lb{ga2}
\end{eqnarray}
Tracing out the distribution
\eqref{600}, we end up with the {Wigner} function  
\begin{align}
\mathcal{W}_{0,0}(x_{1},p_{1}:t)   
  =\frac{1}{\pi(\beta_{2}\eta_{2}-\gamma_{2}^{2})^{\frac{1}{2}}}%
e^{-\Delta_{1}x_{1}^{2}-\Delta_{2}p_{1}^{2}+2\Delta_{12}x_{1}p_{1}}%
\end{align}
and we have
\begin{eqnarray}
&&  \beta_{2}\eta_{2}-\gamma_{2}^{2}=\frac{1}{\tilde{\sigma}_{1}\tilde{\sigma
}_{2}}\left[  \Re(A_{1})\Re(A_{2})+\cos^{2}\phi(t)\sin^{2}\phi(t)\left(
\frac{\dot{h_{1}}}{h_{1}}-\frac{\dot{h_{2}}}{h_{2}}\right)  ^{2}\right]
\\
&&
\Delta_{1}(t,\omega_{c})   =\frac{\tilde{\sigma}_{1}\tilde{\sigma}_{2}%
\eta_{1}}{\left[  \Re(A_{1})\Re(A_{2})+\cos^{2}\phi(t)\sin^{2}\phi(t)\left(
\frac{\dot{h_{1}}}{h_{1}}-\frac{\dot{h_{2}}}{h_{2}}\right)  ^{2}\right]
}\\
&&
\Delta_{2}(t,\omega_{c})    =\frac{\Re(A_{2})}{\left[  \Re(A_{1})\Re(A_{2}%
)+\cos^{2}\phi(t)\sin^{2}\phi(t)\left(  \frac{\dot{h_{1}}}{h_{1}}-\frac
{\dot{h_{2}}}{h_{2}}\right)  ^{2}\right]  }\\
&&
\Delta_{12}(t,\omega_{c})    =-\frac{\tilde{\sigma}_{1}\tilde{\sigma}%
_{2}\gamma_{1}}{\left[  \Re(A_{1})\Re(A_{2})+\cos^{2}(t)\phi\sin^{2}\phi(t)\left(
\frac{\dot{h_{1}}}{h_{1}}-\frac{\dot{h_{2}}}{h_{2}}\right)  ^{2}\right]
}.
\end{eqnarray}
From this result we conclude that the vacuum state $\rho^{A}_{0,0}$ is TDGS
because the {Wigner} function is Gaussian. 
Now
the average of an observable $\mathfrak{B}(x_{1}%
,p_{1})$ can be measured in phase space through $\mathcal{W}_{0,0}(x_{1},p_{1}:t)$,
such as
\begin{equation}
\langle\mathfrak{B}\rangle:=\int dx_{1}dp_{1}\ \mathfrak{B}(x_{1},p_{1})\ \mathcal{W}%
_{0,0}(x_{1},p_{1}:t)%
\end{equation}
which can be used together with 
the identities,  $(J,a)\in\mathbb{R}\times\mathbb{R}^{\ast}_{+}$,
\begin{eqnarray}
&&  \int_{-\infty}^{+\infty}dx\, x e^{-\frac{1}{2}ax^{2}+Jx}=\frac{J}%
{a}\ \left(  \frac{2\pi}{a}\right)  ^{\frac{1}{2}}e^{\frac{J^{2}}{2a}%
}\\
&&  \int_{-\infty}^{+\infty}dx\, x^{2n} e^{-\frac{1}{2}ax^{2}}=\frac
{(2n)!}{a^{n}2^{n}n!}\left(  \frac{2\pi}{a}\right)  ^{\frac{1}{2}}\\
&&  \int_{-\infty}^{+\infty}dx\, x^{2} e^{-\frac{1}{2}ax^{2}+Jx}=\frac{1}%
{a}\ \left(  \frac{2\pi}{a}\right)  ^{\frac{1}{2}}e^{\frac{J^{2}}{2a}%
}\left(  1+\frac{J^{2}}{a}\right)  
\end{eqnarray}
to obtain the average values
\begin{eqnarray}
&&
\langle x_{1}\rangle=\langle p_{1}\rangle=0
\\
&&  \langle x_{1}^{2}\rangle= \frac{\Delta
_{2}}{2\left(\Delta_{1}\Delta_{2}-\Delta_{12}^{2}\right)} \\
&&  \langle p_{1}^{2}\rangle=  \frac{\Delta
_{1}}{2\left(\Delta_{1}\Delta_{2}-\Delta_{12}^{2}\right)}
\end{eqnarray}
showing the uncertainty relations 
\begin{eqnarray}
 &&
\left[\Delta x_{i} \Delta p_{i}\right](t,B)=\frac{1}{2}\left[\frac{1}{(1-S_{L})^{2}}+\gamma_{i}^{2}(t,B) \right]  ^{\frac{1}{2}}\label{us}
\end{eqnarray}
where $\gamma_{i}\equiv \gamma_i(t,B)$ are given in (\ref{ga1}-\ref{ga2}),
$i=1,2$. These relations offer the possibilities to open
some discussions and derive conclusions. Indeed, we notice that
the term  $\frac{1}{2}(1+\gamma_{i}^{2})^{\frac{1}{2}}$ is the lower bound with respect to {\rr Robertson-Schr\"{o}dinger} uncertainty. 
 Now if the marginal state is pure then $S_{L}=0$ and $\gamma_{1}=\gamma_2$,
therefore the uncertainty saturates the lower bound $\frac{1}{2}\left(1+\left(\frac{\dot{h_{1}}h_{1}}{\sigma_{1}(0)}\right)^{2}\right)^\frac{1}{2}$. Moreover, if the oscillations are time independent then the minimality is obtained. In this case  $S_{von}$ and $\mathcal{N}$ vanish meaning that our states are separable, from which we notice that the fluctuations encode quantum correlations between states. When $S_{L}\longrightarrow1$, the fluctuations take an infinite values as well as 
$S_{von}$ and $\mathcal{N}$ become infinite. These cases show in a compact way that the three quantities $(S_{L}, S_{von}, \mathcal{N})$ are connected, which is an important feature.

\section{Results and discussions\label{S7}}

To numerically study {\rr the effect of magnetic field}    on the dynamics of
entanglement, mixedness and uncertainty we use a realistic quenched  model 
\cite{R9,R8}. {\rr In this latter,}  
the 
frequencies and coupling parameter are quenched as  
\begin{equation}
\sigma_{1}(t,\omega_{c})=\left\{
\begin{array}{ll}
(\sigma_{i,1}^{2}+\omega_{c}^{2})^{\frac{1}{2}},& t=0\\
(\sigma_{f,1}^{2}+\omega_{c}^{2})^{\frac{1}{2}},& 0<t\\
\end{array}
\right.,\qquad \sigma_{2}(t,\omega_{c})=\left\{
\begin{array}{ll}
(\sigma_{i,2}^{2}+\omega_{c}^{2})^{\frac{1}{2}},& t=0\\
(\sigma_{f,2}^{2}+\omega_{c}^{2})^{\frac{1}{2}},& 0<t\\
\end{array}
\right.
\end{equation}
\begin{equation}
J(t)=\left\{
\begin{array}{ll}
J_{i},& t=0\\
J_{f},& 0<t\\
\end{array}
\right.
\end{equation}
 where $y=i,f$, $j=1,2$ 
and frequencies $\sigma_{y,j}$ are given by
\begin{eqnarray}
\sigma_{y,1}^{2}&=&\frac{\omega_{y,1}^{2}+\omega_{y,2}^{2}}{2}+\tilde{\kappa} \frac{1}{2}\left(4J_{y}^{2}+(\omega_{y,1}^{2}-\omega_{y,2}^{2})^{2}\right)^{\frac{1}{2}}\\
\sigma_{y,2}^{2}&=&\frac{\omega_{y,1}^{2}+\omega_{y,2}^{2}}{2}-\tilde{\kappa}  \frac{1}{2}\left(4J_{y}^{2}+(\omega_{y,1}^{2}-\omega_{y,2}^{2})^{2}\right)^{\frac{1}{2}} 
\end{eqnarray}
such that $\omega_{y,j}$ are  the quenched values of $\omega_{j}$ 
\begin{equation}
\omega_{j}(t)=\left\{
\begin{array}{ll}
\omega_{i,j},& t=0\\
\omega_{f,j},& 0<t.\\
\end{array}
\right.
\end{equation}
Consequently, the solutions of the {Ermakov} equations  now take the forms
\begin{eqnarray}
h_{1}^{2}(t,\omega_{c}) &=&\frac{\sigma_{f,1}^{2}-\sigma_{i,1}^{2}}{2(\sigma_{f,1}^{2}+\omega_{c}^{2})}\cos\left(2(\sigma_{f,1}^{2}+\omega_{c}^{2})^{\frac{1}{2}}\ t\right)+\frac{\sigma_{f,1}^{2}+\sigma_{i,1}^{2}+2\omega_{c}^{2}}{2(\sigma_{f,1}^{2}+\omega_{c}^{2})}\label{eq82}\\ 
h_{2}^{2}(t,\omega_{c}) &=&\frac{\sigma_{f,2}^{2}-\sigma_{i,2}^{2}}{2(\sigma_{f,2}^{2}+\omega_{c}^{2})}\cos\left(2(\sigma_{f,2}^{2}+\omega_{c}^{2})^{\frac{1}{2}}\ t\right)+\frac{\sigma_{f,2}^{2}+\sigma_{i,2}^{2}+2\omega_{c}^{2}}{2(\sigma_{f,2}^{2}+\omega_{c}^{2})}.\label{eq83}
\end{eqnarray} 
In the next, we inspect the obtained results  to present different plots showing
the behavior of three quantities 
under various choices
of the physical parameters. This will help to understand 
the effect of magnetic field  on the dynamics of our system.

 \subsection{Dynamics of mixedness $S_{L}$}
 
\begin{figure}[htbphtbp]
    \centering
   \includegraphics[width=8.4cm]{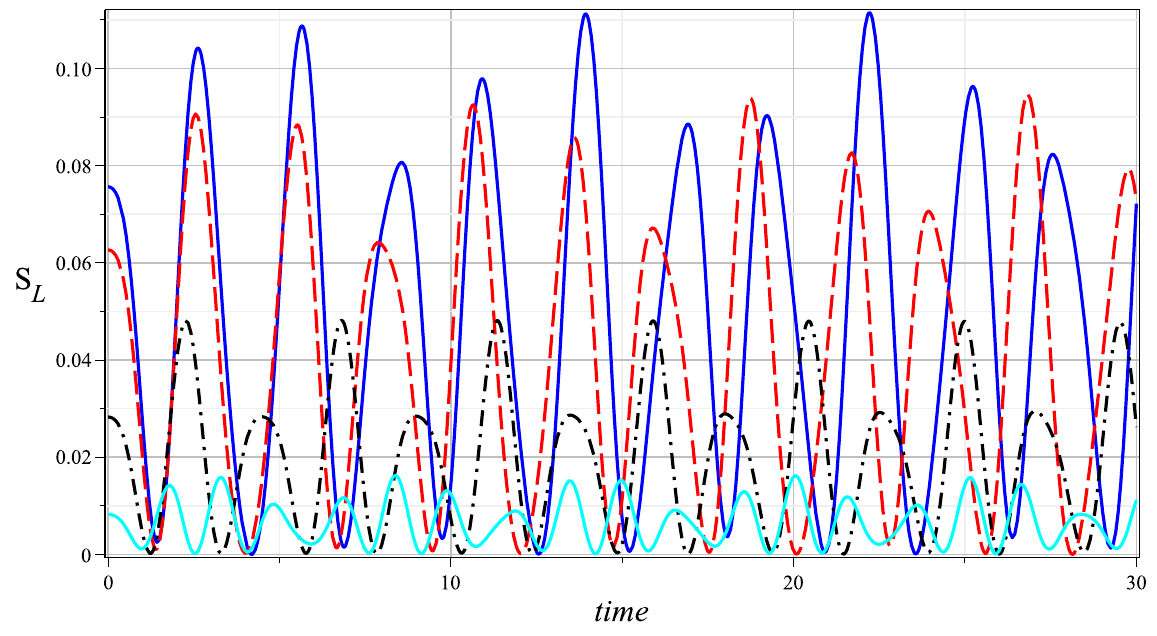}
\captionof{figure}{\sf (color online) 
{\rr The effect of } magnetic field  on the dynamics of mixedness with 
$J_{i}=1.1, J_{f}=0.9, \omega_{i,1}=1, \omega_{f,1}=1.3, \omega_{i,2}=1.5$, $\omega_{f,2}=1.8$, $\omega_{c}=0$ (blue solid line),  $\omega_{c}=0.3$ (red dashed line),  $\omega_{c}=0.8$ (black dashed-dotted line),   $\omega_{c}=1.5$ (cyan solid line).} \label{1f}
\end{figure}

 Remember that the physical meaning of mixedness is the lack of information about the preparation of the state  \cite{R24}. For our Gaussian bipartite vacuum state we have shown that it is symmetric  because $\rho^{A}_{(0,0)}=\rho_{(0,0)}^{B}$, which leads to study the  dynamics of mixedness of one of the both marginal states, for example $\rho_{(0,0)}^{A}$. We use the linear entropy $S_{L}$ as a quantifier of this amount of information or generally one can use also the {Bastiaans-Tsallis} entropies  $S_{BT}^{A}=\frac{1-\Tr(\rho^{\nu})}{\nu-1}$ \cite{R28} and it is worthy to note that $S_{L}= S_{BT}^{A}$ $(\nu=2)$. To show the  effect of  magnetic field  $\left(\omega_{c}=\frac{eB}{2c}\right)$ 
 on the dynamics of mixedness we plot in \textsf{Figure} \ref{1f} $S_{L}$ versus time under the quench $(\omega_{i,1}=1, \omega_{i,2}=1.5, j_{i}=1.1)\longrightarrow (\omega_{f,1}=1.3,\omega_{f,2}=1.8,J_{f}=0.9)$. One can see that when $\omega_{c}=0$ $(B=0)$  the amount of mixedness exhibits a bi-sinusoidal behavior in the  time scale $[0,30]$. 
 While for $\omega_c\neq 0$,  we observe that those oscillations undergo an 
 amplitude frequency modulation, 
 which decreases and then the amount of mixing decreases. Indeed, a large $\omega_{c}$  yields to the oscillations in isotropic regime  then we have $A_{12}\longrightarrow0$ and $S_{L}\longrightarrow0 $, see Eq.~\eqref{38}. The small bi-oscillations are due to  solutions  $h_{i}$ of the {Ermakov} equations and their time derivatives $\dot{h}_{i}$,
 $i=1,2$. The increasing in multi-frequencies is due to  the phase $\sim\omega_{c}t$ in the both $h_{i}$ and $\dot{h}_{i}$. We conclude that magnetic field purifies our TDGS and the mixedness of marginal states  can  be driven by a magnetic field. 
 
\begin{figure}[htbphtbp]
    \centering
   \includegraphics[width=8.4cm]{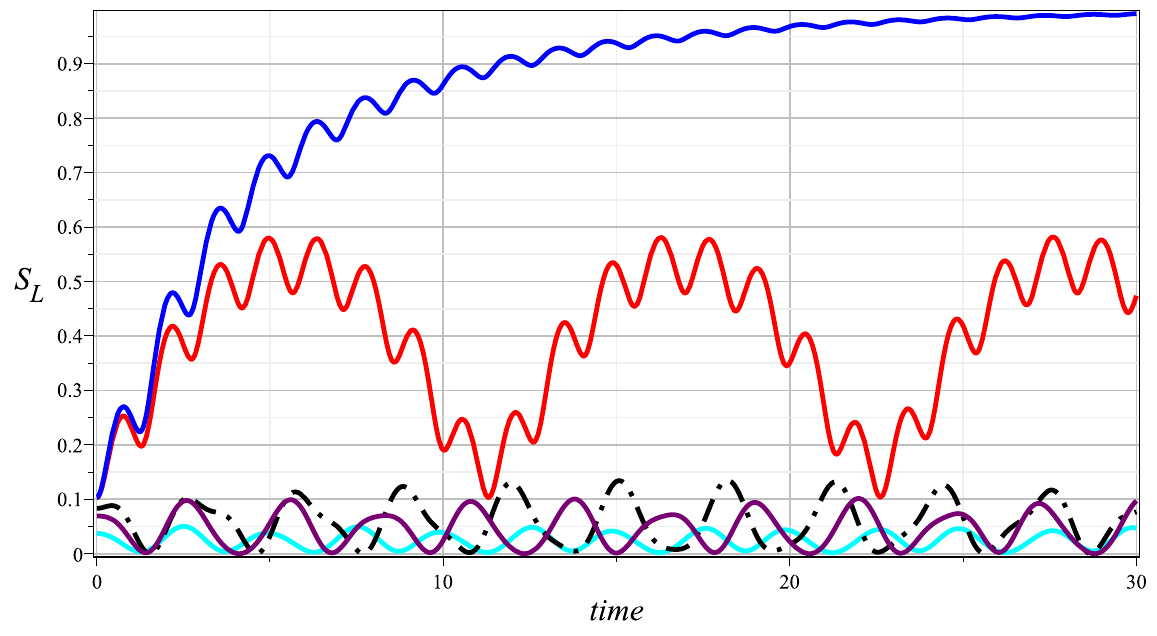}
   \captionof{figure}{\sf (color online)
{\rr The effect of} quenched coupling parameter $J_{f}$  on the dynamics of mixedness with $J_{i}=1.1,  \omega_{i,1}=1$, $\omega_{f,1}=1.3$  $\omega_{i,2}=1.5$, $\omega_{f,2}=1.8$, $\omega_{c}=0.2$,  $J_{f}=0.5$ (cyan solid line),  $J_{f}=0.9$ (purple solid line),  $J_{f}=1.2$ (black dotted-dashed line), $J_{f}=2.3$ (red solid line), $J_{f}=2.4$ (blue solid line).}\label{2f}
\end{figure}

 The quenched value of the coupling parameter $J_{f}$ plays 
 an interesting role in the dynamics of mixedness. In \textsf{Figure} \ref{2f} we show that when $\omega_{c}=0.2$, $J_{i}=1.1$ and $(\omega_{i,1}=1, \omega_{i,2}=1.5)\longrightarrow (\omega_{f,1}=1.3,\omega_{f,2}=1.8)$ the marginal states will be more mixed as the coupling parameter increases. For $J_{f}\leq 0.5$  the correlations between two subsystems are very weak and the mixedness presents a tiny  sinusoidal oscillations behavior, which is due to the fact that a weak coupling $J\longrightarrow0$ yields to  $\phi\longrightarrow 0$ and eventually $S_{L}\longrightarrow0$. Increasing $J_{f}$, the  dynamics of $S_{L}$ periodic oscillations  exhibit an increasing of amplitude and decreasing of frequency.  For a large coupling these oscillations   take an exponential dynamics, which is due to the nonphysical oscillations of the first oscillator (i.e. $\sigma_{1}\in \mathbb{C}$), then trigonometric functions $(\cos, \sin)$ in $h_{1}$ transform as $(\cosh,\sinh)$.
 But it is important to notice that this effect can be removed with a suitable choice of $\omega_{c}$ that allow  us to  construct  exponential  solutions
 of {Ermakov} equations \cite{R27} in order  to follow the dynamics of  time dependent 
 harmonic oscillators. \\

\begin{figure}[htbphtbp]
    \centering
   \includegraphics[width=8.4cm]{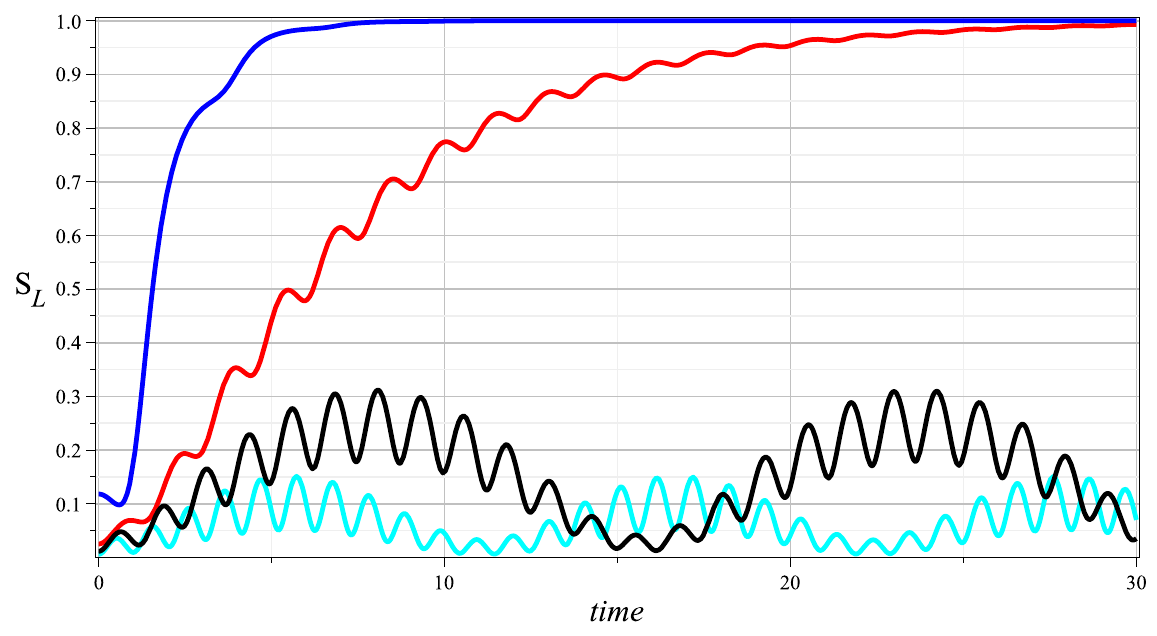}
\captionof{figure}{\sf (color online) {\rr The effect of}
quenched coupling frequency $\omega_{f,2}$  on the dynamics of mixedness with $J_{i}=1.1,  J_{f}=0.9,  \omega_{i,1}=1, \omega_{i,2}=1.5$, $\omega_{f,1}=0.4$, $\omega_{c}=0.1$,  $\omega_{f,2}=3$ (cyan solid line), $\omega_{f,2}=2.5$ (black solid line), $\omega_{f,2}=2$ (red solid line), $\omega_{f,2}=0.5$ (blue solid line).} \label{3f} 
\end{figure}

In \textsf{Figure} \ref{3f} we present the effect of 
quenched coupling frequency $\omega_{f,2}$ on the dynamics of mixedness. We observe that the mixedness undergoes an amplitude frequency modulation, 
decreases as the quenched frequency increases and eventually the oscillations disappear while the dynamics becomes exponential. Finally, it appears that the magnetic field plays an important role in purification of marginal states and gives  rise to the physical oscillations with high correlations. We observe that a large coupling yields quickly to maximally mixed states and the small values of the  quenched frequencies lead to maximally  marginal mixed states.  

\subsection{Dynamics of phase space fluctuations}

\begin{figure}[htbphtbp]
    \centering
   \includegraphics[width=8.4cm]{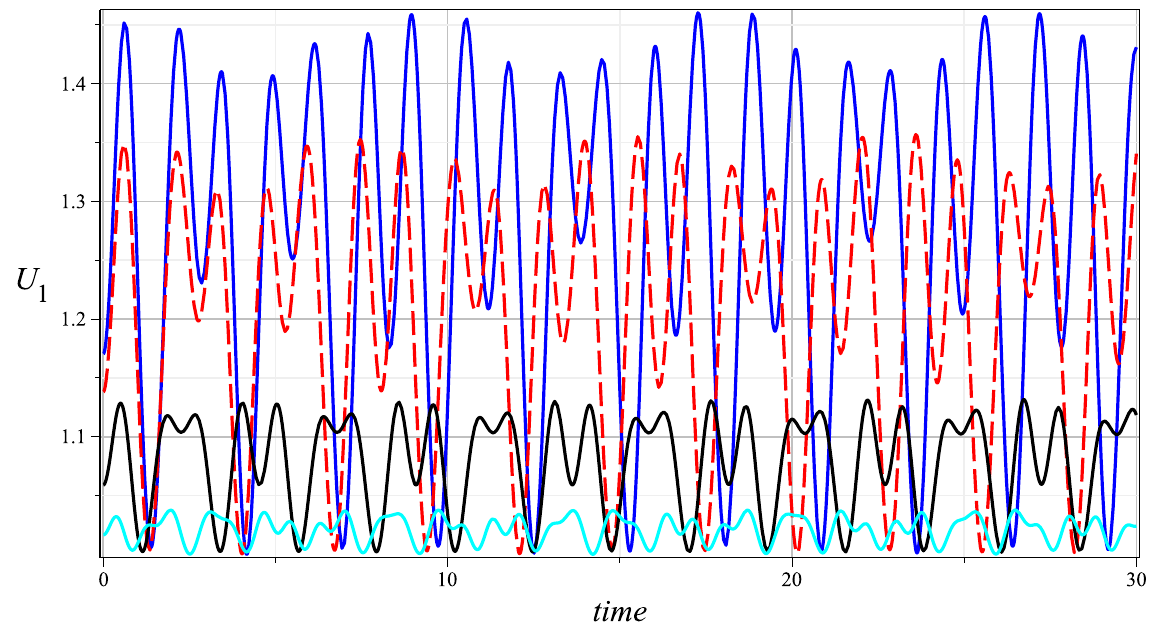}
\captionof{figure}{\sf(color online) {\rr The effect of}
magnetic field on the dynamics of uncertainty $U_{1}=(2\Delta x_{1}\Delta p_{1})^{2}$ with $J_{i}=1.1, J_{f}=0.9, \omega_{i,1}=1, \omega_{f,1}=1.3, \omega_{i,2}=1.5$, $\omega_{f,2}=1.8$,  $\omega_{c}=0$ (blue solid line),  $\omega_{c}=0.3$ (red dashed line),  $\omega_{c}=0.8$ (black solid line),  $\omega_{c}=1.5$ (cyan solid line).} \label{1fu}
\end{figure}

 To  graphically study the dynamics of uncertainty, we plot the variation of $U_1=\left(2\Delta x_{1}\Delta x_{2}\right)^{2}$ versus time under suitable conditions. Firstly, we investigate {\rr the effect of} magnetic field on such dynamics in \textsf{Figure} \ref{1fu} by taking a fix quench $(\omega_{i,1}=1, \omega_{i,2}=1.5, J_{i}=1.1)\longrightarrow (\omega_{f,1}=1.3,\omega_{f,2}=1.8, J_{f}=0.9)$ and different values of $\omega_{c}=0,0.3,0.8,1.5$. For $\omega_{c}=0$ the dynamics of uncertainty presents a multi-oscillatory behavior, which is due to solutions of the {Ermakov} equations and their derivatives. 
 Increasing $\omega_{c}$ undergoes the dynamics to oscillate in the vicinity of minimality. We conclude from our analysis that uncertainty can be assisted via a static magnetic field and eventually handle the quantum fluctuations.  \\
 
\begin{figure}[htbphtbp]
    \centering
   \includegraphics[width=8.4cm]{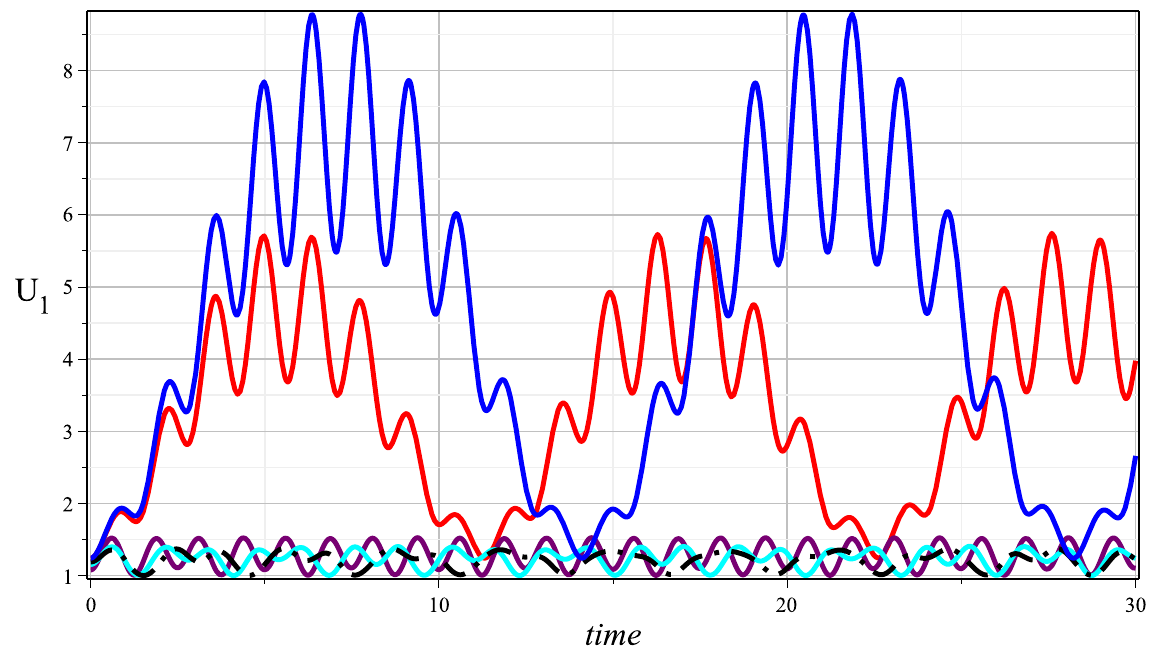}
\captionof{figure}{\sf (color online) {\rr The effect of}
magnetic field  on the dynamics of uncertainty $U_{1}=(2\Delta x_{1}\Delta p_{1})^{2}$ with $J_{i}=1.1, \omega_{i,1}=1, \omega_{f,1}=1.3, \omega_{i,2}=1.5$, $\omega_{f,2}=1.8$,  $\omega_{c}=0.2$,  $J_{f}=0.5$ (purple solid line), $J_{f}=0.9$ (cyan solid line),  $J_{f}=1.2$ (black dotted-dashed line),  $J_{f}=2.3$ (red solid line),   $J_{f}=2.33$ (blue solid line).} \label{2fu}
\end{figure}

 Secondly, we show the impact of the quenched coupling parameter $J_{f}$ on the dynamics of uncertainty $U_{1}=(2\Delta x_{1}\Delta p_{1})^{2}$  in \textsf{Figure} \ref{2fu}. 
 It is clearly seen that for small values of the coupling, the dynamics of uncertainty exhibits a periodic behavior. However,  for a large coupling such dynamics  exhibits a bi-sinusoidal behavior with a large amplitude and small frequency. From the critical value $J_{f}=2.4$ the oscillations disappear and the behavior become exponential, which  is quite natural because as we have seen previously for mixedness, $J_{f}=2.4$ yields  to
 $S_{L}\longrightarrow 1$ and from Eq. (\ref{us}) the first term goes to infinity, i.e. $\frac{1}{(1-S_{L})^{2}}\longrightarrow+\infty$. Note that, the lower bound of uncertainty  $\frac{1}{2}(1+\gamma_{1}^{2}(t,\omega_{c}))^{\frac{1}{2}} $ strongly depends on the magnetic field and time,
 which is very important to preserve the invariance of uncertainty with respect to phase space  transformations during the action of the magnetic field and  dynamics. Such lower bound  is the same as that obtained using the {\rr Robertson-Schr\"{o}dinger} uncertainty \cite{R32}.\\

\begin{figure}[htbphtbp]
    \centering
   \includegraphics[width=8.4cm]{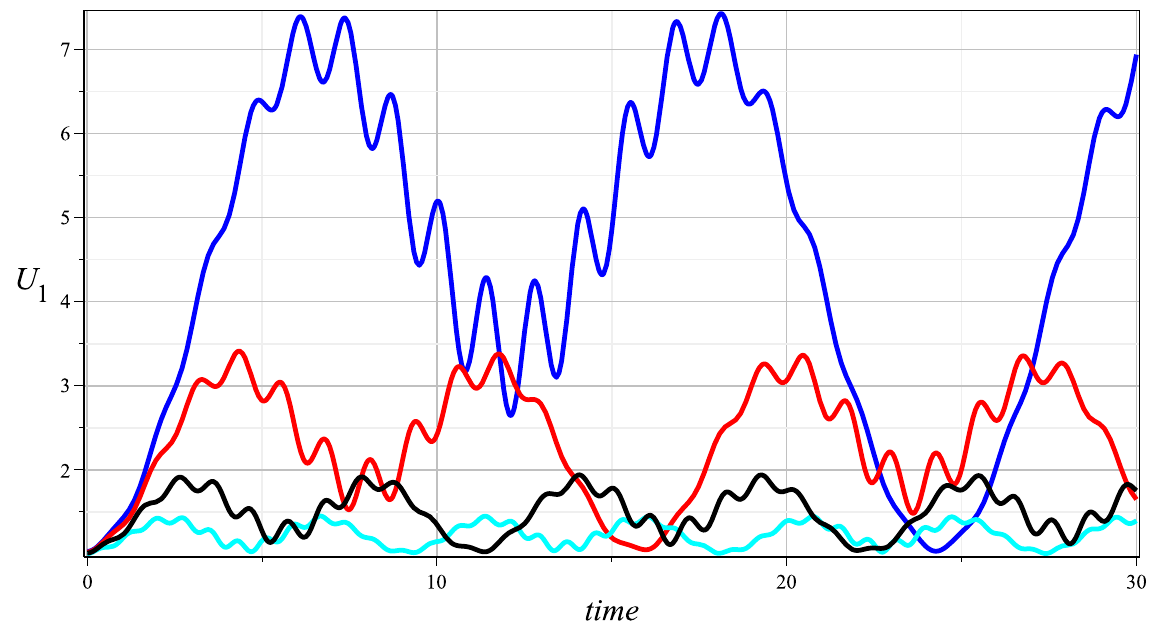}
\captionof{figure}{\sf(color online)
The effect of quenched coupling frequency $\omega_{f,2}$ on the dynamics of uncertainty $U_{1}=(2\Delta x_{1}\Delta p_{1})^{2}$ with $J_{i}=1.1,  J_{f}=0.9,  \omega_{i,1}=1,\omega_{i,2}=1.5$, $\omega_{f,1}=0.4$, $\omega_{c}=0.1$,  $\omega_{f,2}=4$ (cyan solid line), $\omega_{f,2}=3$ (black solid line), $\omega_{f,2}=2.5$ (red solid line), $\omega_{f,2}=2.3$ (blue solid line).} \label{3fu} 
\end{figure}

 Thirdly, we show the effects of quenched frequency $\omega_{f,2}$ on the dynamics of uncertainty $U_{1}=(2\Delta x_{1}\Delta p_{1})^{2}$  in \textsf{Figure} \ref{3fu} 
 for different quenches $(\omega_{i,1}=1, \omega_{i,2}=1.5, J_{i}=1.1)\longrightarrow (\omega_{f,1}=0.4, \omega_{f,2}=2.3,2.5,3,4, J_{f}=0.9)$. We observe that
 for  $\omega_{f,2}=0.5,2$, the uncertainty presents a large uncertainty with exponential behavior, which is trivial because the dynamics drives $S_{L}$ to $1$ then $U_{1}\longrightarrow+\infty$.  It is interesting to notice  that a large $\omega_{f,2}$ yields  to the pure marginals  and the lower bound will be saturated. 
 We conclude that the fluctuations depend on the magnetic field, coupling parameter and  quenched frequencies. When the mixedness is minimal, the uncertainty is also 
 because a tiny  mixedness means that the lack of information is very small then the fluctuations are minimal.\\

 \subsection{Dynamics of entanglement via $S_{von}$ and $\mathcal{N}$}
 
 Our global state is pure and Gaussian, thus to study the dynamics of  entanglement we use two quantifiers. First one is the {von Neumann} Entropy or generally the {R\'{e}nyi} Entropies $S_{R}^{\nu}(\rho^{A}_{(0,0)})=\frac{1}{1-\nu}\ln \Tr\left[\left(\rho_{(0,0)}^{A}\right)^{\nu}\right]$, where 
 $S_{von}=\lim_{\nu \rightarrow 1}S_{R}^{\nu}$ \cite{R8,R24},  and second is the logarithmic negativity $\mathcal{N}$ \cite{R1,R24}. The later quantifier presents a lot of simplifications because its  measure does not involve the resolution of  spectral equations Eq. (\ref{esp}) and addresses only to the marginal purities and some symplectic parameters of CM. Moreover, in our case TDGE is symmetric then $\mathcal{N}$  will be expressed only in term of marginal purity $\Tr\left[\left(\rho_{(0,0)}^{A} \right)^{2}\right]$. To  graphically show the dynamics of the amount of entanglement encoded in the reduced states $\left(\rho^{A}_{(0,0)},\rho^{B}_{(0,0)}\right)$ we plot both quantities $S_{von}$ and $\mathcal{N}$ versus time.
 
\begin{figure}[htbphtbp]
    \centering
   \includegraphics[width=8.2cm]{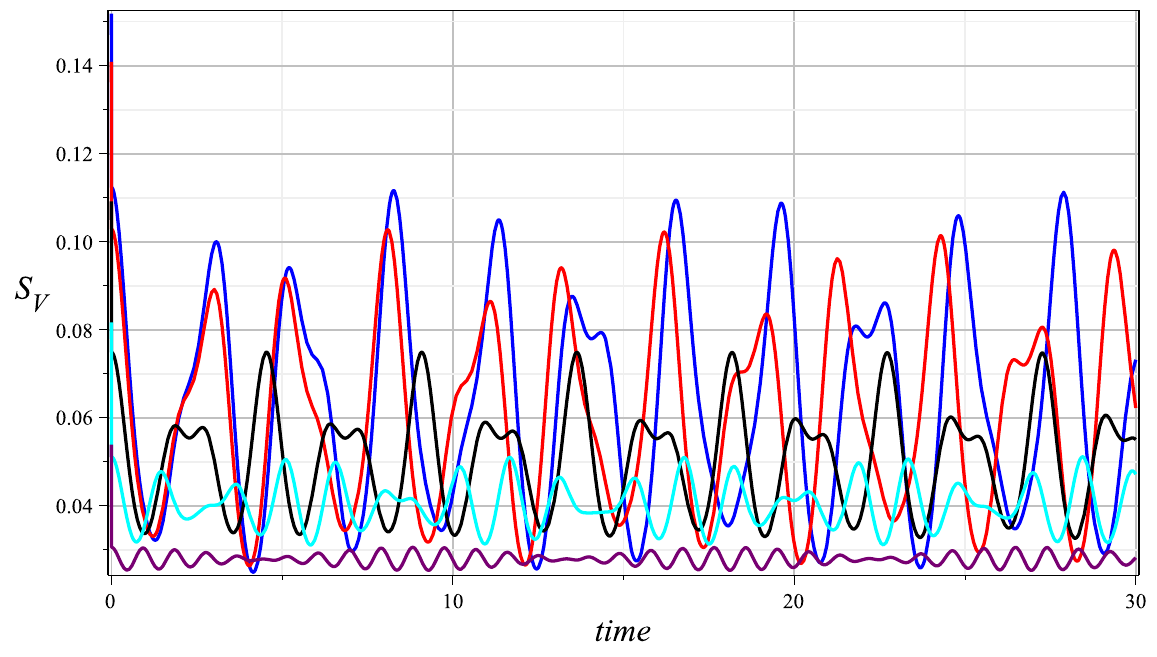}
   \includegraphics[width=8.4cm]{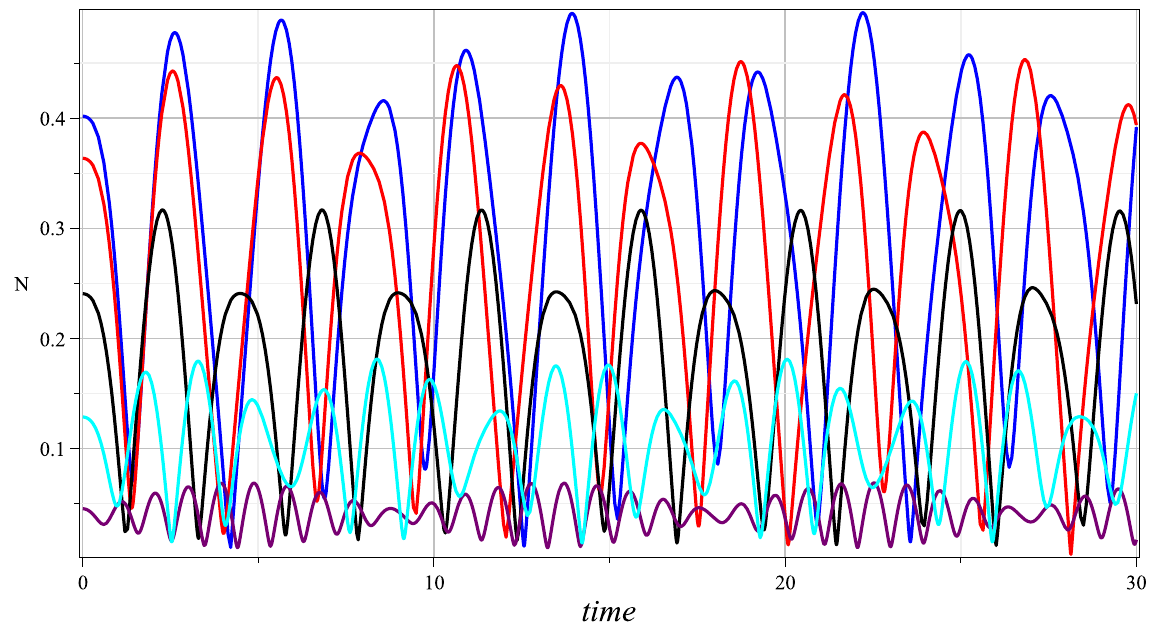}
\captionof{figure}{\sf (color online) {\rr The effect of}
magnetic field  on the dynamics of {von Neumann} entropy $S_{von}$ and logarithmic negativity $\mathcal{N}$ with $J_{i}=1.1,\, J_{f}=0.9, \omega_{i,1}=1,\omega_{f,1}=1.3,\,\omega_{i,2}=1.5$,   $\omega_{f,2}=1.8$,   $\omega_{c}=0$ (blue solid line), $\omega_{c}=0.3$ (red solid line), $\omega_{c}=0.8$ {{(black solid line)}},  $\omega_{c}=1.5$ (cyan solid line), $\omega_{c}=3$ {{(purple solid line)}}. 
} \label{1fs}
\end{figure}

 Firstly, we investigate the effect  produced by a magnetic field 
 in \textsf{Figure} \ref{1fs} 
 by taking fix quenches $(\omega_{i,1}=1,\omega_{i,2}=1.5, J_{i}=1.1)\longrightarrow (\omega_{f,1}=1.3,\omega_{f,2}=1.8, J_{f}=0.9)$ and different values of $\omega_{c}=0,0.3,0.8,1.5,3$. For $\omega_{c}=0$,  
 $S_{von}$ and $\mathcal{N}$ exhibit a similar topological behavior {{(the same variations)}}. For $\omega_{c}=0.3,0.8,1.5,1.3$, the magnetic field creates an important decreasing on the amplitude  of $S_{von}$  and $\mathcal{N}$. Thus the entanglement can be remoted by the magnetic field and $\omega_{c}$ affects violently the frequency  of  both oscillations,
 which is natural because of the phase $\sim \omega_{c}t$ presented in solutions of the {Ermakov} equations. This tell us the magnetic field can be used to
 control the information delivered by $S_{von}$  and $\mathcal{N}$. \\
 
\begin{figure}[htbphtbp]
    \centering
   \includegraphics[width=8.4cm]{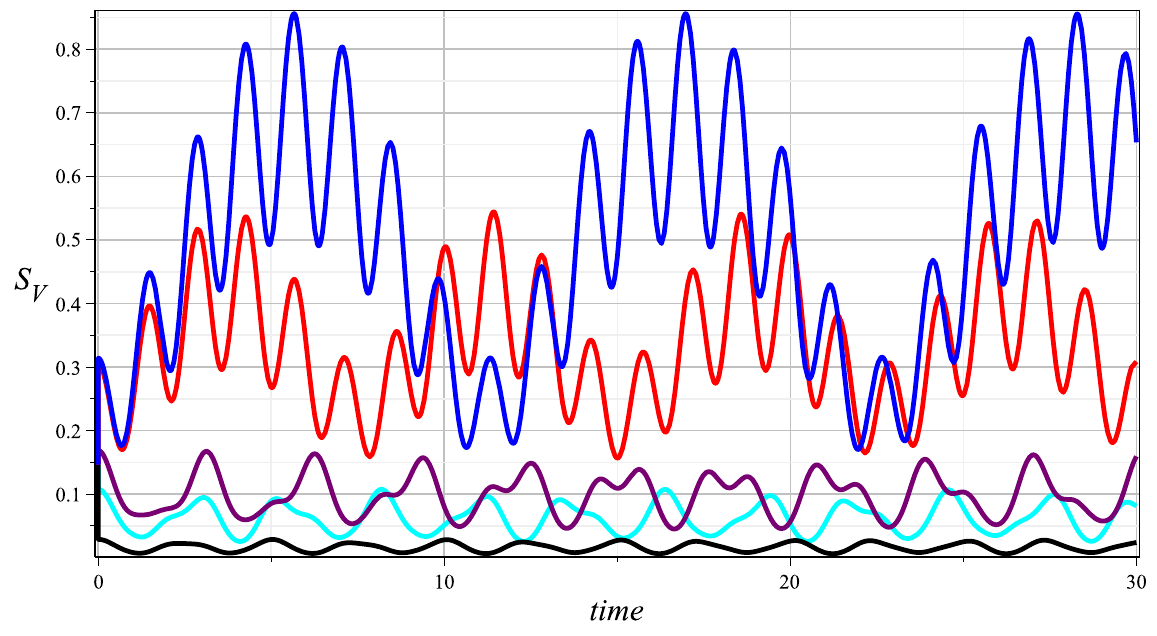}
   \includegraphics[width=8.4cm]{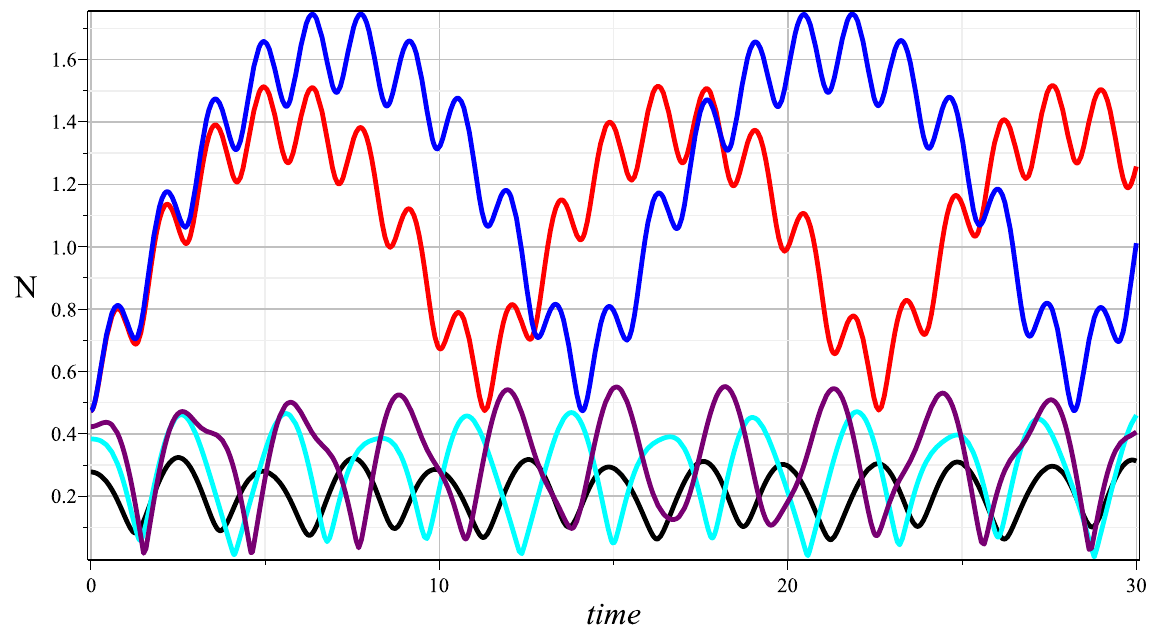}
\captionof{figure}{\sf (color online) {\rr The effect of}
quenched coupling $J_{f}$ on the dynamics of {von Neumann} $S_{von}$ entropy and
logarithmic negativity $\mathcal{N}$ with $J_{i}=1.1, \omega_{i,1}=1,\,\omega_{f,1}=1.3,\omega_{i,2}=1.5$, $\omega_{f,2}=1.8$,   $\omega_{c}=0.2$,  $J_{f}=0.5$ (black solid line), $J_{f}=0.9$ (cyan solid line),  $J_{f}=1.2$ (purple solid line),  $J_{f}=2.3$ (red solid line),  $J_{f}=2.33$ (blue solid line).} \label{2fs}
\end{figure}

 Secondly, we explore the impact of the quenched coupling $J_{f}$ on the dynamics of entanglement in \textsf{Figure} \ref{2fs}. 
 Note that,  if $J=0$  then $\phi=0$ and $S_{von}=\mathcal{N}=0$ showing that  the coupling $J_{f}$  witnesses the  existence of entanglement. Now  increasing $J_{f}$ to $0.5$ and $0.9$,  the entanglement dynamics undergoes an amplitude frequency modulation: the amount of entanglement  becomes more important and  exhibits a bi-sinusoidal behavior, which due to solutions of the {Ermakov} equations. We observe that
a large coupling $J_{f}$ yields to the nonphysical oscillations (negativity of the square of frequency) $\sigma_{1}\in\mathbb{C}$,  which leads to $S_{L}\rightarrow +\infty$ then $\mathcal{N},S_{von}\longrightarrow +\infty$. 

\begin{figure}[htbphtbp]
    \centering
   \includegraphics[width=8.4cm]{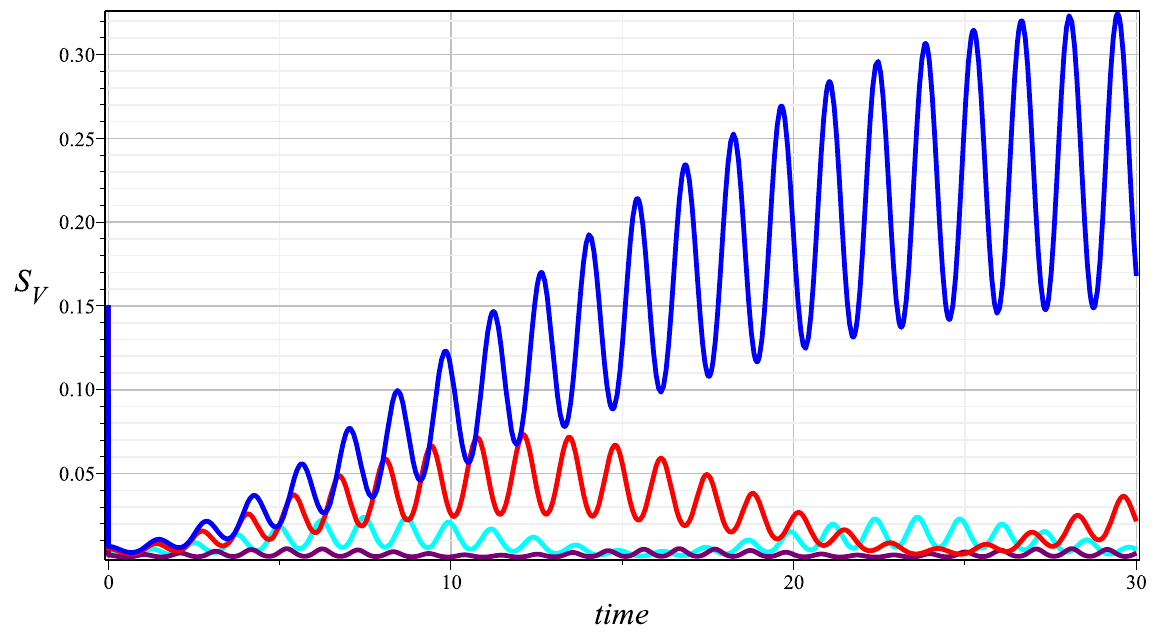}
   \includegraphics[width=8.4cm]{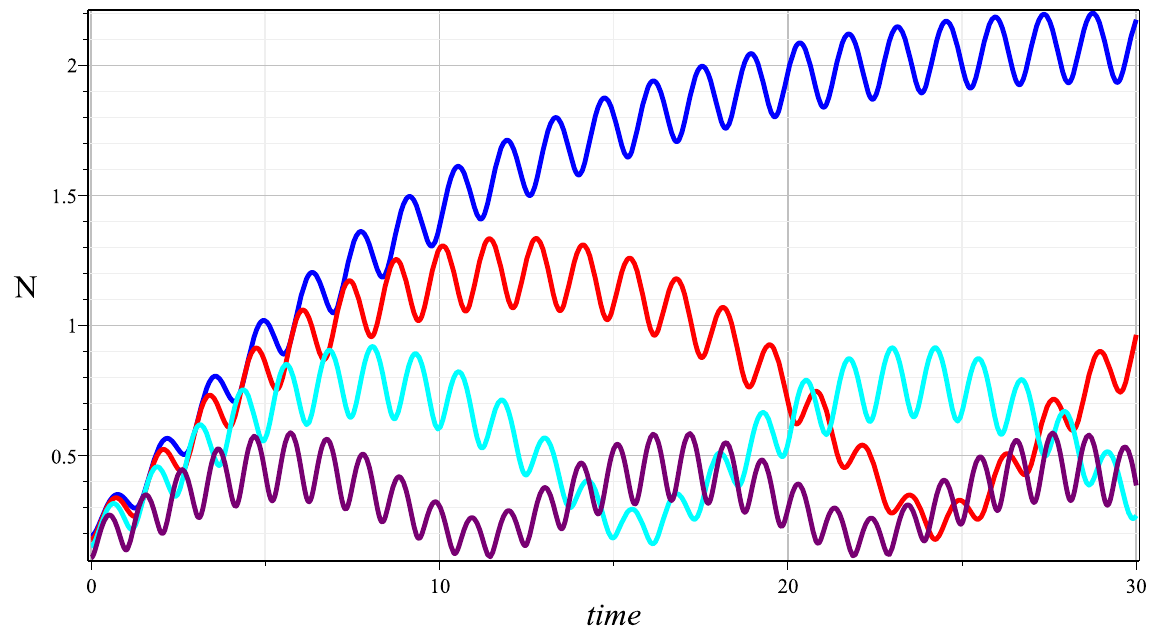}
\captionof{figure}{\sf (color online)
{\rr The effect of}  quenched frequency $\omega_{f,2}$ on the dynamics of {von Neumann} $S_{von}$ entropy and
logarithmic negativity $\mathcal{N}$ with $J_{i}=1.1, \omega_{i,1}=1, \omega_{f,1}=1.3,\omega_{i,2}=1.5$, $\omega_{f,2}=1.8$,  $\omega_{c}=0.1$,  $\omega_{f,2}=3$ (purple solid line),  $\omega_{f,2}=2.5$ (cyan solid line),  $\omega_{f,2}=2.3$ (red solid line),
 $\omega_{f,2}=2.2$ (blue solid line).} \label{3fs}
\end{figure}

Thirdly, we show the  variations of $S_{von }$ and $\mathcal{N}$  versus time with respect to quenched frequency $\omega_{f,2}$ (without loss  of generality) in \textsf{Figure} \ref{3fs}. 
The dynamics shows that if we increase $\omega_{f,2}$ the amount of entanglement decreases  because if 
$\omega_{f,2}$ increases then the difference $\vert \omega_{f,1}^{2}-\omega_{f,2}^{2}\vert$ becomes large implying that the separability is reached.  Moreover, two entangled microscillators should have  the {{same}} mechanical features. It is interesting to notice  that the three dynamics are linked and present the {{same}} topological behavior with respect to $\omega_{c}, J_{f}$, $\omega_{f,2}$ and similar dynamics.\\

\section{Conclusion\label{S8}}

We have considered time dependent harmonic oscillator subject to
a static magnetic field
and studied the dynamics of entanglement, mixedness and logarithmic negativity.
%
 Firstly, we have used the rotation $SO(2)$ in the phase plane $(x,p)$ to diagonalize the Hamiltonian.
 We have derived the solutions $\Psi_{n,m}$ of TDSE of our system
 and focused only on the vacuum state $\Psi_{0,0}$ by showing that it is TDGS, symmetric and pure state. This was used to show that the {von Neumann} entropy $S_{von}$  and logarithmic negativity $\mathcal{N}$ are legitimate quantifiers of entanglement.
 Furthermore, we have  computed the common marginal purity and further the linear entropy $S_{L}$ to quantify the degree of mixedness. In addition, we have employed
 the {Heisenberg} uncertainty to study the dynamics of uncertainties and eventually demonstrated  explicit relations between the entanglement, 
  mixedness and uncertainty, which allowed us to use the purity or mixedness as suitable candidates of the required quantifiers. 

Subsequently,
we have studied the dynamics of the entanglement, mixedness and uncertainty  using the quenched model \cite{R9,R8} in order to derive the  solutions $h_{i}$ of the {Ermakov} equations and their time derivatives $\dot h_i$, $i=1,2$.  We have shown that the magnetic field purifies the marginal states thus decreasing the amount of quantum correlations, a feature that can be used to control and handle  these prototypical states. We have demonstrated also that the three quantities (entanglement, mixedness, logarithmic negativity)  present the same behavior with respect to the magnetic field, which leads to construct a  meaningful quantifier mixedness-based. We have shown also that the uncertainty approaches to saturate that is lower bound in the vicinity of separable states $\mathcal{N},S_{von}\rightarrow 0$, an issue used to detect experimentally the entangled states.  


\section*{Acknowledgment}

The generous support provided by the Saudi Center for Theoretical
Physics (SCTP) is highly appreciated by AJ.

\end{document}